\documentstyle[12pt,graphicx,color,subfigure,epsfig,cite,amsmath,lineno]{article}
\def\baselinestretch{1.3}
\newcommand{\ba}{\begin{array}}
\newcommand{\ea}{\end{array}}
\newcommand{\bd}{\begin{displaymath}}
\newcommand{\ed}{\end{displaymath}}
\newcommand{\be}{\begin{equation}}
\newcommand{\ee}{\end{equation}}
\newcommand{\bea}{\begin{eqnarray}}
\newcommand{\eea}{\end{eqnarray}}

\def\fb{\, {\rm fb}}
\def\met{E_T \hspace*{-1.1em}/\hspace*{0.5em}}
\def\tev{\, {\rm TeV}}
\def\gev{\, {\rm GeV}}
\def \wt{\widetilde}
\def \lsptwo{\wt\chi_2^0}
\def \lspone{\wt\chi_1^0}
\def \chonem {{\wt\chi_1^\pm}}
\def \chargino2 {{\wt\chi_2^\pm}}
\def \lstop{\wt{t}_{1}}
\def \ch2m {{\wt\chi_2^-}}
\def \lspone{\wt\chi_1^0}
\def \lspi{\wt\chi_i^0}
\def \chonep {{\wt\chi_1^+}}
\def \charginoi {{\wt\chi_i^\pm}}

\def\a{\alpha}

\def\q2 {q^2}

\def\bt{\begin{table}}
\def\et{\end{table}}

\catcode`@=11 
\def \gsim{\mathrel{\mathpalette\@versim>}}
\def \lsim{\mathrel{\mathpalette\@versim<}}
\def \@versim#1#2{\lower0.4ex\vbox{\baselineskip\z@skip\lineskip\z@skip
     \lineskiplimit\z@\ialign{$\m@th#1\hfil##\hfil$%
     \crcr#2\crcr\sim\crcr}}}
\catcode`@=12 

\def\mhalf{m_{1/2}}

\def\wt{\widetilde}
\def\te{\tilde e}

\def\tu{\tilde u}

\def\tb{\tilde b}

\def\tst{\tilde t}

\def\tg{\tilde g}

\def \lspi{\wt\chi_i^0}

\def \lspone{\wt\chi_1^0}
\def \mlspone{m_{\lspone}}
\def \lsptwo{\wt\chi_2^0}
\def \mlsptwo{m_{\lsptwo}}
\def \lspthree{\wt\chi_3^0}
\def \mlspthree{m_{\lspthree}}
\def \lspfour{\wt\chi_4^0}
\def \mlspfour{m_{\lspfour}}

\def\issue(#1,#2,#3){{\bf #1}, #2 (#3)}

\def\PREP(#1,#2,#3){Phys.\ Rep. \issue(#1,#2,#3)}

\begin{document}

\begin{flushright}
{ HRI-RECAPP-2009-014}
\end{flushright}

\begin{center}

{\large\bf Non-universal scalar mass scenario with Higgs funnel region of 
SUSY dark matter: a signal-based analysis for the Large Hadron Collider}\\
[15mm] Subhaditya Bhattacharya$^{a}$\footnote{subha@hri.res.in}, 
Utpal Chattopadhyay$^{b}$\footnote{tpuc@iacs.res.in}, 
Debajyoti Choudhury$^{c}$\footnote{debchou@physics.du.ac.in}, \\[1ex]
Debottam Das$^{b}$\footnote{tpdd@iacs.res.in} 
and Biswarup Mukhopadhyaya$^{a}$\footnote{biswarup@mri.ernet.in}\\[4ex]

{\it $^{(a)}$Regional Centre for Accelerator-based Particle Physics,\\
Harish-Chandra Research Institute,\\
Chhatnag Road, Jhusi, Allahabad 211 019, India.}\\[1ex]
    {\it $^{(b)}$Department of Theoretical Physics, 
    Indian Association for the Cultivation of Science, \\2A \& 2B 
Raja S.C. Mullick Road, Kolkata 700 032, 
India. }\\[1ex]
{\it $^{(c)}$Department of Physics and Astrophysics,
University of Delhi, Delhi 110 007, India}.
\end{center}

\begin{abstract} 
  We perform a multilepton channel analysis in the context of the Large Hadron
  Collider (LHC) for Wilkinson Microwave Anisotropy probe (WMAP) compatible 
points in a model with non-universal
  scalar masses, which admits a Higgs funnel region of supersymmetry dark 
matter 
  even for a small $\tan\beta$.
  In addition to two and three-lepton final states, four-lepton 
  events, too, are shown to be useful for this purpose.
  We also compare the collider signatures in similar channels for WMAP
  compatible points in the minimal supergravity (mSUGRA) framework with
  similar gluino masses.  Some definite features of such non-universal
  scenario emerge from the analysis.
\end{abstract}

\vskip 1 true cm

\newpage
\setcounter{footnote}{0}

\def\baselinestretch{1.5}
\section{Introduction} 
Low energy Supersymmetry (SUSY)\cite{SUSY} is a strong candidate for physics
beyond the Standard Model. A general framework like the Minimal
Supersymmetric Standard Model (MSSM)\cite{SUSY,KaneKingRev}, however, suffers
from a large number of 
{\em a priori} 
unrelated parameters. This 
lack of predictability
can be minimized only if one assumes a definite
mechanism for breaking SUSY. The minimal supergravity (mSUGRA) \cite{msugra}
model that assumes gravity to be the mediator between the hidden sector
wherein supersymmetry breaks and the observable sector (where the MSSM exists)
is relatively economical in this respect. Starting from only a few parameters,
renormalization group evolutions (RGE)
 and the use of radiative electroweak symmetry
breaking (REWSB) generate all the MSSM parameters at the electroweak scale.

The mSUGRA scenario has a remarkable simplicity of principle, an economy of
parameters and features that at least partially ameliorate potentially
disastrous consequences in low energy physics.  From a more agnostic
standpoint, however, there is no strong reason to restrict ourselves to such
universal models. 
For one, even with gravity conveying supersymmetry
breaking, the soft SUSY-breaking terms need not be universal at the 
supergravity scale,
but would depend on the structure of the K\"ahler potential. Similarly, 
large non-universal corrections may accrue to the soft parameters as 
a result of the evolution between the Planck scale and the 
gauge-coupling unification scale~($M_G\simeq 2\times 10^{16}$~GeV) 
\cite{Polonsky:1994rz}. These and other
related issues have led to several studies 
of non-universal scalar
\cite{ucddOct2008,berez,Nath:1997qm,Cerdeno:2004zj,ellis-all,baer-all-non,
baer-higgs,so101,so102,Datta:1999uh,BM-SB-AD2} and gaugino mass
\cite{nonunigaugino,BM-AD-SB} models. 
Non-universal scalar masses may appear due to a non-flat
K\"ahler metric\cite{soni}, or, for example, from $SO(10)~D$-terms 
\cite{so101,so102,Datta:1999uh}. 
However,
any such nonuniversality, at the electroweak scale, would lead to low-energy
flavor changing neutral current (FCNC) processes 
(through SUSY loops)\cite{FCNC}. The
existing data on flavor physics thus impose severe constraints on any
nonuniversality in scalar masses, in particular for the first two families.
The restrictions on the third generation scalars (and the Higgses) from FCNC
data are not too severe though.

It turns out that both FCNC and CP-violation constraints may be best tackled
if one assumes the first two generations of scalars to be multi-TeV and
(quasi-)degenerate in masses~\cite{Gabbiani:1996hi}\footnote{We remind the 
reader
that satisfying constraints imposed by electric dipole moments of electron and
neutron would require very large scalar masses if we like to have finite
values for the CP-violating SUSY phases.}. Clearly, allowing universal scalar
masses at the gauge coupling unification scale would not satisfy the above
objective because either $(i)$ the REWSB constraint would prohibit such large
scalar masses for a reasonable set of values of the gluino masses, or $(ii)$
one must have very large gaugino masses, 
so as to allow very large scalar
masses, thereby worsening the fine tuning problem~\cite{BG}. We recall that,
within
 the MSSM, the naturalness problem and its solution revolve around the third
family, as well as the gaugino and Higgs 
scalar mass parameters.  As long as the third generation
scalars and the electroweak gauginos are on the lighter side, any quantitative
measure of naturalness would stay within an acceptable domain.
Furthermore, constraints from FCNC and CP-violation 
are relatively weak 
in such a scenario with an inverted 
mass hierarchy\cite{Barger-inverted,RIMHall}. 

In this work, we consider a particular non-universal scalar mass scenario
(NUSM), namely that of Ref.{\cite{ucddOct2008}}. The model 
addresses the FCNC issue by invoking very large masses for
the first two generations of squarks and sleptons. 
As is well-known, such a solution is difficult to achieve within the mSUGRA
scenario as the requirement of REWSB prevents the scalar masses from being too
large. In the present context, this is circumvented by allowing the third
generation squark masses and the Higgs scalar mass parameters 
to be small. This very smallness also serves to keep the degree of fine-tuning 
within control.

As far as the third generation sleptons are concerned, a very small
SUSY-breaking mass at the GUT scale is not phenomenologically viable since the
larger Yukawa coupling serves to drive down the mass of the lighter stau, 
thereby rendering it the lightest of the supersymmetric partners (LSP) at the
electroweak scale. Consequently, the SUSY-breaking mass in this sector has to
be sizable\footnote{However, in 
analyses with Higgs-exempt no-scale SUSY model\cite{wells} or 
in a model with gaugino mediation\cite{Buchmuller}
one may avoid such charged LSPs at the electro-weak scale by 
using non-zero Higgs scalar masses
at the unification scale. In these scenarios the no scale boundary 
conditions are also valid for sleptons.}. Rather than introducing a new parameter, we shall assume it to be
same as that of first two generations of squarks or sleptons. To summarize,
at the GUT scale, all sfermion masses are diagonal; and, apart from those
pertaining to the stop and the sbottom, are universal. The 
last-mentioned,
along with the Higgs scalars, have a vanishing mass at this scale.  
While this
construction might seem artificial, note that this accords a special status
to only those fields that 
are expected to play a direct role in EWSB.
Interestingly, the model satisfies the WMAP constraint\cite{WMAPdata} 
on neutralino  
relic density
for a large
region of the parameter 
space without requiring any delicate mixing of Binos and
Higgsinos. For simplicity, we confine ourselves to a universal 
gaugino mass and a
vanishing trilinear soft-breaking parameter ($A_0$) 
at $M_G$.

We investigate how such  a scenario
can leave its fingerprint on numbers measured at the Large Hadron
Collider (LHC). Such fingerprints are of value if ways
can be devised to distinguish this scenario from an mSUGRA one
with, say, similar gluino masses. For this, one has to 
perform a multichannel analysis \cite{baer11,BM-AD-SB,BM-SB-AD2,ad-uc-lhc} 
studying  several final states
simultaneously. 

A promising signal of supersymmetry (with a conserved R-parity)
comprises large missing transverse energy, accompanied 
by  
jets and leptons with varying multiplicities. An analysis in different 
channels, compared with that of a similar mSUGRA scenario may lead to a 
significant hint of the non-universality.
In the present analysis, we assess
the accessibility of our non-universal scalar mass model (NUSM)  
at the LHC. We find that the
direct pair production of stops and sbottoms 
as well as their cascading down from gluino decays lead to
the possibility of four-lepton final states as a distinct signature
of this scenario. Additionally, we also analyse 
the two-lepton and the three-lepton final states. This includes 
opposite sign dilepton, same-sign dilepton and trilepton final states.
All these analyses are done also for mSUGRA so that the multipronged 
approach of analysing for different channels may become more 
conclusive.  

   The paper is organized as follows. In Section~\ref{nusmAndConstraints} we 
describe the NUSM model, apply cosmological constraints on 
neutralino dark matter and use low energy constraints such as those 
from $b \rightarrow s +\gamma$ or $B_s \rightarrow \mu^+ \mu^-$. We also 
identify benchmark points for our analyses of collider signals at the LHC. 
In Section~\ref{collidersection}, we pinpoint our strategies for collider 
simulations and report the numerical results. Finally, in 
Section~\ref{conclusionsection}, we summarize our results and conclude. 

\section{The Non-Universal Scalar Mass model (NUSM) and benchmark points}
\label{nusmAndConstraints}
\subsection{The NUSM parameter space}
\label{nusmpara}
The NUSM model \cite{ucddOct2008}, at the scale $M_G$, 
is characterized by five parameters, namely,\
\begin{equation}
\tan\beta,
\,  \mhalf, \, m_0, \, A_0 {\rm ~and~} sign(\mu) \ .
        \label{eq:param_defn}
\end{equation}
The parameters, here, play r\^oles similar to those in mSUGRA
except for a subtle and important difference in the scalar sector.
Masses of the first two generations of scalars (squarks and sleptons) and 
the third generation of sleptons are assigned the value
$m_0$. However, the
Higgs scalars and the third family of squarks have 
vanishing mass values at $M_G$. Here, $m_0$ is allowed to be up to 
tens of TeVs. 
As has already been stated, we limit ourselves to a vanishing $A_0$ in this 
analysis. We have considered $\mu>0$ in this analysis. 

The NUSM admits 
a  smaller pseudoscalar Higgs boson 
mass $m_A$ on account of the  
Large Slepton Mass (LSM) renormalization group effect\cite{ucddOct2008} 
for large $m_0$.
With such a $m_0$, the LSM effect 
causes $m_{H_D}^2$ to become large and 
negative and this may happen for even a small 
$\tan\beta$. 
This, in turn,  
reduces the masses of the pseudoscalar Higgs boson ($A$), 
the CP-even heavy Higgs boson ($H$) and the charged Higgs bosons ($H^\pm$).  
In this scenario, $\mu$ is quite insensitive
to a change in $m_0$\cite{ucddOct2008},
since the Higgs and the third-generation squark 
masses at $M_G$ are free of the latter. In fact, 
$\mu$ is completely independent of $m_0$ up to one-loop, 
whereas
the two-loop contributions to its RGEs result in only 
a tiny  dependence on $m_0$.
Recall that, in mSUGRA 
on the contrary, $|\mu|$ decreases significantly with 
an increase in $m_0$.
Whereas this led to a very small 
$|\mu|$ for a large $m_0$ in mSUGRA, giving the so called 
Hyperbolic Branch/Focus point (HB/FP)\cite{hyper,focus} region that is close to the upper limit 
of $m_0$ satisfying REWSB for a given 
$\mhalf$, there is no HB/FP type of  effect in NUSM and $\mu$ stays 
reasonably independent of $m_0$. 
It turns out that the lightest neutralino 
is highly Bino-dominated (with a small Higgsino admixture)
throughout virtually the entire
parameter space of NUSM. 
Along with the resonance condition, namely, 
$2\,m_{\tilde \chi_1^0} \simeq m_A \, (m_H)$, 
 the small Higgsino content allows 
the LSP to have  the right degree of pair-annihilation 
via $s$-channel Higgs-exchanges, 
so as to satisfy the WMAP limits on
the neutralino relic density.
It is important to note
that, excepting for LSP-stau coannihilation, 
the Higgs-pole annihilation 
mechanism is the only one
in NUSM that reduces 
the relic density from overabundance to an 
acceptable degree of abundance. 
Thus, unlike in models such as the mSUGRA, here one does not need any 
delicate mixing between a Bino and Higgsinos in order to satisfy 
the WMAP data. Such Higgs-pole annihilations 
that occur for large $\tan\beta$ in mSUGRA is 
typically known as the funnel region\cite{dreesDM93,funnel}. 
NUSM has an extended funnel region 
that spans from low to high $\tan\beta$.  

\subsection{Cosmological and low energy constraints in NUSM}
\label{CosmoAndLowEn}
\noindent
Assuming that dark matter was generated thermally, 
the limits on the cosmological relic density from the 
WMAP data\cite{WMAPdata} impose severe 
constraints on supergravity type of models 
wherein 
the lightest neutralino $\lspone$ becomes the LSP for 
most of the parameter space\cite{DMreview,recentSUSYDMreview}. 
We now perform an analogous analysis for the NUSM.
For a given set of parameter (vide eq.(\ref{eq:param_defn})) values,
 the supersymmetric particle 
spectrum is generated using {\tt SuSpect} v2.34\cite{suspect}. 
This, then, is used as an input 
to 
{\tt micrOMEGAs}\cite{micromegas} for computing the 
neutralino relic density. 
The recent WMAP
data\cite{WMAPdata} stipulates that, at the 3$\sigma$ level,
\begin{equation}
0.091 < \Omega_{CDM}h^2 < 0.128 \ ,
\label{relicdensity}
\end{equation}
where $\Omega_{CDM}$ is the dark matter 
relic density in units of the critical
density and $h=0.71\pm0.026$ is the reduced Hubble constant 
(namely, in units of
$100 \ \rm km \ \rm s^{-1}\ \rm Mpc^{-1}$).

\begin{figure}[!hb]

\centerline{ \psfig{file=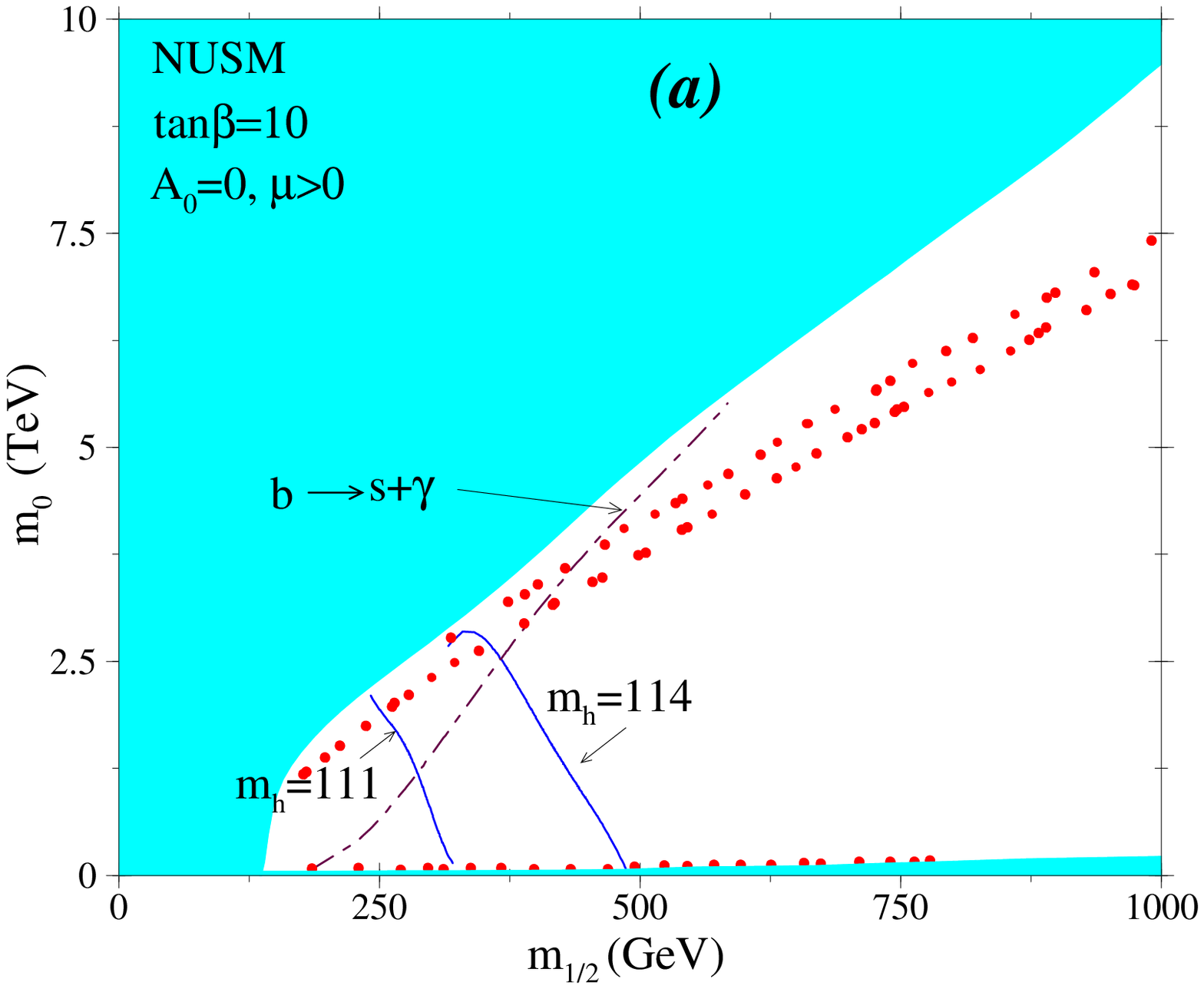,width=6. cm,height=6cm,angle=0}
\hskip -15pt \psfig{file=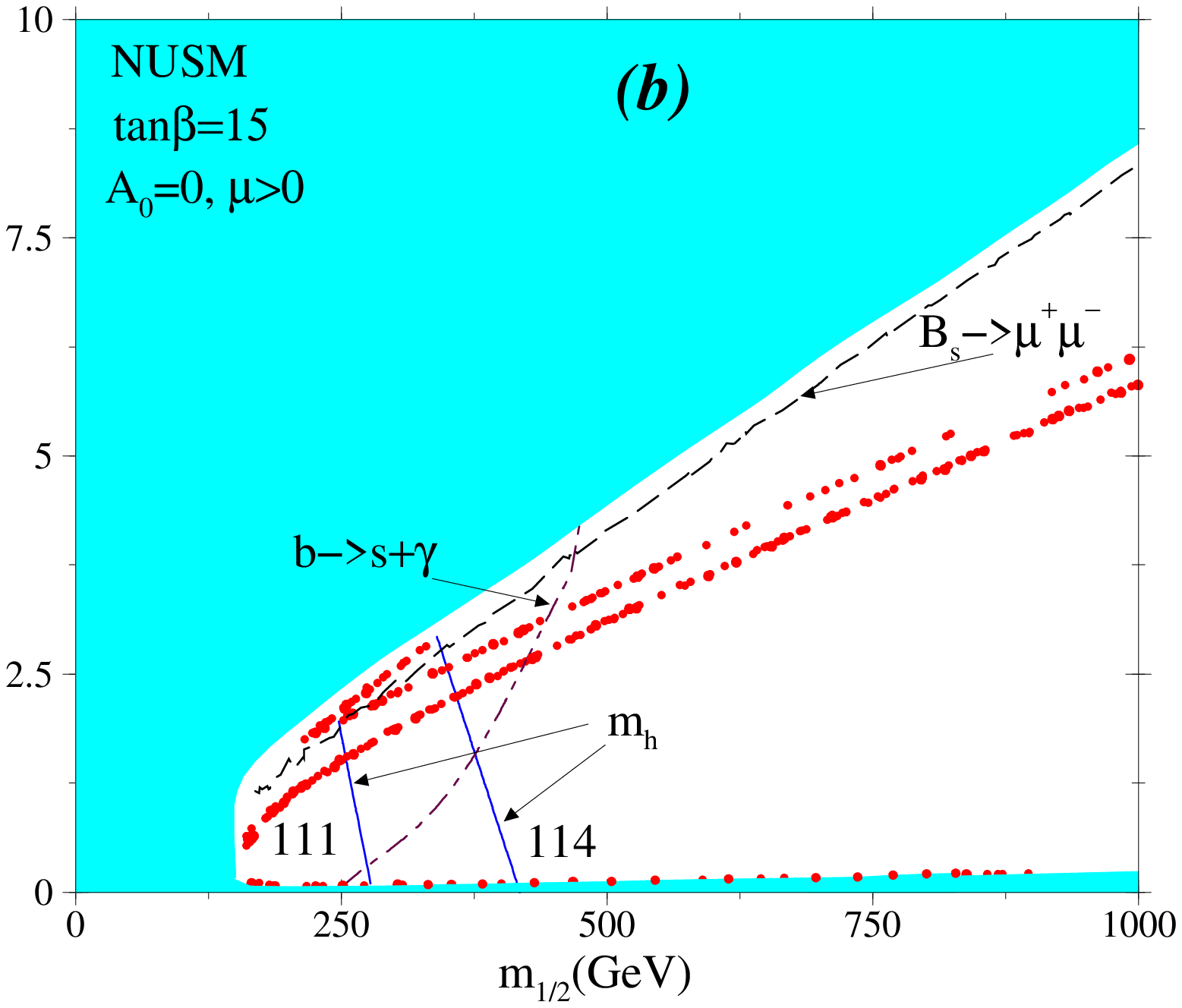,width=6. cm,height=6cm,angle=0}
\hskip -15pt \psfig{file=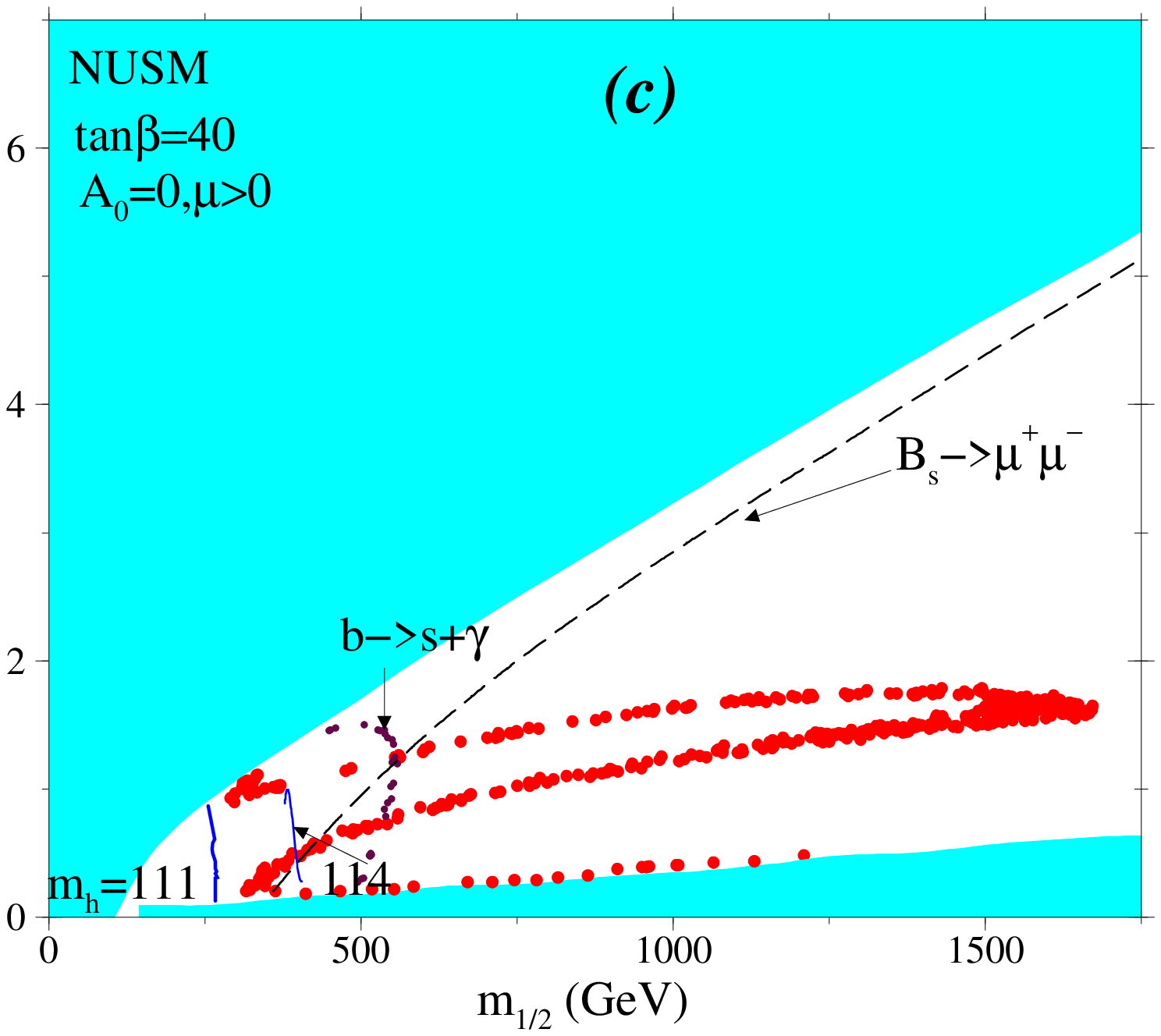,width=6. cm,height=6cm,angle=0}}
\caption{\em 
~(a) WMAP allowed regions in the $\mhalf-m_0$ plane for $\tan\beta=10$ 
and $A_0=0$ with $\mu>0$ for NUSM are shown in red dots. 
Lighter Higgs boson mass limits are represented by solid lines. 
Dot-dashed line refers to $b\rightarrow s \gamma$ limit. The 
entire region is allowed by $B_s\rightarrow \mu^+ \mu^-$ data. 
~(b) Same as (a) except that $\tan\beta=15$. The 
$B_s\rightarrow \mu^+ \mu^-$ bound is shown as a  
long-dashed line.  This eliminates a small strip of region below 
the discarded top (cyan) region.  
~(c)Same as (b) except that $\tan\beta=40$.
}
\label{mhalfmzerodm}
\end{figure}

In Fig.\ref{mhalfmzerodm}, we display the allowed regions in the
$\mhalf-m_0$ plane for three values of the ratio of the Higgs vacuum
expectation values, namely $\tan\beta=10,15$ and $40$.  The thin 
(cyan)
sliver at the bottom is ruled out as, for such values of the parameters, the
lighter stau becomes the LSP.  The upper (cyan) region is
rejected primarily on account of the failure in the breaking of the
electroweak symmetry via radiative means. In other words, for such parameter
values, $m_A^2$ does not acquire a 
positive value through RG flow. Close to
the boundary of this region, several other phenomenological constraints become
important.  The most important of these pertain to $(i)$ the LEP2 
and Tevatron lower bounds
for sparticle masses, $(ii)$ sfermions turning tachyonic, or
$(iii)$ the appearance of charge and color breaking (CCB) minima. To be
allowed, a parameter point must evade all these and other such constraints.
Specific details may be found in Ref.\cite{ucddOct2008}.

Highlighted (in bold---red---dots) in Fig.\ref{mhalfmzerodm} 
are examples of parameter points that satisfy the 
WMAP data. There are two distinct regions with acceptable relic density
as already mentioned in Sec.{\ref{nusmpara}}. 
$(a)$ The Higgs pole annihilation region (also known as the 
funnel region) is characterised by   
$2m_{\tilde \chi_1^0} \simeq m_A, m_H$. In this particular 
scenario, it extends over the full range of $\mhalf$ under consideration. 
The Higgs pole annihilations may occur through $s$-channel 
pseudoscalar Higgs boson ($A$) or CP-even neutral H or h-bosons. 
NUSM has a bino-dominated LSP similar to what occurs in mSUGRA 
in its funnel region that satisfies the WMAP data.   
Similar to the case in mSUGRA, the WMAP satisfied parameter 
regions of NUSM is also dominantly characterized by the pseudoscalar 
Higgs boson mediated resonance annihilation.   
The exact or near-exact 
resonance regions have very large annihilation cross sections resulting in  
a high degree of under-abundance of dark matter. The resonance 
region that satisfies the WMAP data may be a few $\Gamma_{A/H}$ 
away from exact resonance.  The widths $\Gamma_{A/H}$ (of $A/H$ bosons)  
can be fairly large ({\em e.g.} $\Gamma_{A/H} \sim 10$-$50$~GeV).
The WMAP satisfied regions fall on either side of the exact 
resonance condition thus showing two branches in the figure.   
($b$) The second region, just above the lower ruled-out part,
corresponds to the case where the lighter stau is nearly degenerate 
with the LSP, leading to very efficient LSP-stau coannihilation, thereby 
reducing the relic abundance to acceptable levels.

Also imposed on Fig.\ref{mhalfmzerodm} are the pertinent low-energy
constraints. Whereas non-observance at LEP2 impose a strict bound of 114.4
GeV on the SM Higgs\cite{hlim}, with recent negative results from
Tevatron\cite{higgs_tevatron}
ruling out even somewhat heavier Higgses, 
the translation of this bound to the MSSM case
needs careful consideration. Apart from the parameter-dependence of the
cross-sections at LEP2/Tevatron, one needs to account for the uncertainties in
computing the mass of light Higgs boson\cite{higgsuncertainty}, originating
primarily from momentum-independent as well as momentum-dependent two-loop
corrections, higher loop corrections from the top-stop sector
etc. Numerically, this amounts to about 3 GeV, and we have taken that into
account in drawing the solid lines representing this constraint. 
Additionally, a part of the NUSM parameter space is associated with  
very light $m_A$ even for a small $\tan\beta$ 
and this may lower the lighter Higgs boson lower bound 
to a value much smaller than that of the SM Higgs boson limit. We will 
revert to this while discussing the NUSM benchmark points.

A low energy observable of particular importance is the decay rate for 
$b \rightarrow s \gamma $ rate\cite{bsg-sm,bsg-bsm}, which, at the 
$3 \sigma$ level, reads\cite{bsg-recent}
\begin{equation}
2.77 \times 10^{-4} < Br (b \rightarrow s \gamma) < 4.33 \times 10^{-4}.
\label{bsgammalimits}
\end{equation}
We used {\tt micrOMEGAs}\cite{micromegas} for computation of 
$b \rightarrow s \gamma $ that in turn refers to Refs.\cite{bsg-sm,bsg-bsm} 
for actual computation.    
Typically, $b \rightarrow s \gamma $ disfavours the 
small $\mhalf$ region where the rate is below 
the lower limit.  Note, however, that the usual estimation assumes a
perfect alignment at high energies between the quark and squark mass
matrices. In other words, the (super-)Cabibbo-Kobayashi-Maskawa matrix
operative for supersymmetric diagrams is assumed to be identical to the usual
CKM matrix. 
However, if one relaxes the above assumption 
and considers even a 
moderate amount of $\tilde b-\tilde s$ mixing 
at the GUT scale, Eq.\ref{bsgammalimits} is no longer
an effective constraint for high scale models like mSUGRA.  
This, on the other hand, will not cause any significant change in the
sparticle mass spectra or in the flavor conserving process of neutralino
annihilation.
We refer the reader to Refs.\cite{djouadidmsugra,bsg-okumura} 
for further discussions on the amount of model-dependence in computing 
$Br(b \rightarrow s \gamma)$ in this context. 

Since the NUSM scenario may contain a light pseudoscalar Higgs, it is 
necessary to consider
the constraints from $B_s\rightarrow \mu^+ \mu^-$. Within the MSSM, the
above branching ratio is proportional to $m_A^{-4}$ and
${{\tan}^6\beta}$\cite{bsmumurefs}.  The recent CDF\cite{CDF} limit for
$Br(B_s \to \mu^+ \mu^-)$ is given by
\begin{eqnarray}
{\rm Br} ( B_s \to \mu^+ \mu^-) < 5.8 \times 10^{-8}.
\label{Bsmumu}
\end{eqnarray}
The branching ratio of $B_s\rightarrow \mu^+ \mu^-$ is 
evaluated by using {\tt micrOMEGAs}\cite{micromegas} that in turn 
implemented Ref.\cite{Bobeth:2001sq} for the computation. 
The computation involves 
inclusion of loop contributions due to chargino, sneutrino, stop and 
Higgs exchanges.   
The upper limit of this branching ratio is shown in dashed lines in
Fig.\ref{mhalfmzerodm}.  The white regions above the dashed lines in
Figs.\ref{mhalfmzerodm}$(b,c)$ are thus discarded. 
As mentioned in Ref.\cite{ucddOct2008}, the
intense coupling region of Higgs bosons that appears when $m_A$ is very small
is also ruled out in NUSM for the same reason.

\subsection{A few benchmark points}
\label{benchmarksubsection}
The NUSM has a large volume of allowed
parameter space, especially because REWSB 
does not prohibit $m_0$ 
from assuming a very large value. We focus here 
on a few characteristic parameter points that satisfy WMAP 
as well as low 
energy constraints. As seen in Figs.\ref{mhalfmzerodm}({\em a--c}), 
for 
a given $\mhalf$, 
the upper limit on $m_0$ decreases with an increase in $\tan\beta$. 
As an example, for $\mhalf=1$~TeV, $m_0$ may well be as large as 
$7$~TeV for $\tan\beta=10$, $6$~TeV for $\tan\beta=15$, and
$1.6$~TeV for $\tan\beta=40$. 
\begin{table}[!ht]
\begin{center}
\begin{tabular}[ht]{|c|c|c|c|}
\hline
parameter & A & B & C \\
\hline
$\tan\beta$ &10  &15 &40\\
$\mhalf$ &270  &255 &540\\
$m_0$ &2050  &2000 &1250\\
$A_0$ & 0  & 0 & 0\\
$sign(\mu)$ &1  &1 &1\\
\hline
$\mu$ & 312 & 291 &  651  \\
$m_{\tg}$ & 709 &   674 & 1280   \\
$m_{\tu_L}$ & 2100  &  2050  & 1660  \\
$m_{\tst_1}$ &  276 & 248 & 842  \\
$m_{\tst_2}$ &   493 & 465 & 1030  \\
$m_{\tb_1}$ & 390 & 354 &  958   \\
$m_{\tb_2}$ &  434 & 403 & 1020  \\
$m_{\te_L}$ & 2050 & 2000  &1300   \\
$m_{{\tilde \tau}_1}$ & 2040  & 1970  & 1120   \\
$m_{{\tilde\chi_1}^{\pm}}$ & 196 &183  & 430  \\
$m_{{\tilde\chi_2}^{\pm}}$ & 347 & 327 & 668 \\
$\mlspfour$ & 347 & 326 & 668  \\
$\mlspthree$ & 318 & 297 & 655  \\ 
$\mlsptwo$ & 197 & 185&  430 \\ 
$\mlspone$ & 108 & 101 & 226  \\ 
$m_A$ & 259 &  148 &  403  \\
$m_{H^+}$ &  272 &  169 &  411  \\
$m_h$ & 111 & 111 &116 \\
$\Omega_{{\tilde \chi}_1^0}h^2$& 0.105 & 0.102 & 0.130 \\
$Br(b\to s\gamma)$ & $1.59\times 10^{-4}$ & $4.65\times 10^{-5}$ 
& $2.73\times 10^{-4}$ \\
$Br(B_s\to \mu^+\mu^-)$ & $4.02\times 10^{-9}$ & $2.81\times 10^{-8}$ 
& $5.29\times 10^{-8}$ \\
$\Delta a_\mu    $ & $9.31 \times  10^{-11}$ & $1.59 \times  10^{-10}$ 
& $6.99 \times  10^{-10}$ \\
\hline
\end{tabular}
\end{center}
\caption{\em NUSM Benchmark points A, B and C (masses are in GeVs). The 
first five parameters define the model, while the rest are predictions.}
\label{tabnusm}
\end{table}

Here, we have preferentially explored those regions in the parameter space 
which give distinctly different low energy and cosmological 
signatures as compared to mSUGRA. As we have mentioned before, the 
Higgs funnel region for mSUGRA is found only for large values of 
$\tan\beta$. The NUSM 
is characteristically different from mSUGRA in the sense that 
funnel regions exist even for small $\tan\beta$.
Hence, we choose to explore one benchmark point 
with small $\tan\beta$. Additionally, we examine a benchmark point 
for large $\tan\beta$. The next point to note is that 
NUSM, typically, has heavier spectra for 
the first two generations of scalars and the 
third 
generation of sleptons. Sfermions become heavier with 
increase in $\mhalf$. Therefore, if we like to probe the model in the early 
phase of the LHC, we would rather select $\mhalf$ to be relatively small for 
the benchmark points. 

Table~\ref{tabnusm} lists three benchmark points for 
NUSM. Point A (for $\tan\beta=10,A_0=0,\mhalf=270{\rm~GeV},
m_0=2.05{\rm~TeV} {\rm~and~} sign(\mu)=1$),
is associated 
with reasonably 
small masses for stop, sbottom, charginos as well as neutralinos. 
Furthermore, it has 
a light Higgs spectrum. All these are promising from the viewpoint of 
early LHC results. 
Point B of Table~\ref{tabnusm} refers to a special
parameter point ($\tan\beta=15,A_0=0,\mhalf=
255{\rm~GeV}$, $m_0=2.0{\rm~TeV} 
{\rm~and~} sign(\mu)=1$), for which the Higgs sector
is not in the decoupling\cite{decouplingHiggs,nondecouplingHiggs} 
region. Thus, here we obtain a reduced lower limit for $m_h$ 
(close to $M_Z$). 
Point C ($\tan\beta=40,A_0=0,\mhalf=540{\rm~GeV}$, $          
m_0=1.25{\rm~TeV} {\rm~and~} sign(\mu)=1$) 
represents a relatively
heavier spectrum.
However, the relevant parts of NUSM spectra still 
remain within the LHC reach.
We point out that we have relaxed the $b \rightarrow s\gamma$ constraint 
for points A and B. This is in keeping with the discussion in the 
paragraph following
Eq.\ref{bsgammalimits}. However, with a small displacement of 
the parameter point, we would be able to respect the constraint 
at the cost of having a benchmark point with an upwardly shifted spectrum. 
Both points A and B obey the constraints from  
$B_s\rightarrow \mu^+ \mu^-$. Point C, on the other hand, 
satisfies all the constraints, namely, those from $B_s\rightarrow \mu^+ \mu^-$ 
and $b \rightarrow s\gamma$, over and above those from the WMAP data. 

\begin{table}[!ht]
\begin{center}
\begin{tabular}[ht]{|c|c|c|c|}
\hline
parameter & mSUGRA-A & mSUGRA-B & mSUGRA-C \\
\hline
$\tan\beta$ &10  &15 &40\\
$\mhalf$ &253  &252 &490\\
$m_0$ &2740  &2300  & 2680 \\
$A_0$ & 0  & 0 & 0\\
$sign(\mu)$ &1  &1 &1\\
\hline
$\mu$ & 139 & 135 &  266  \\
$m_{\tg}$ & 740 &   725 & 1270   \\
$m_{\tu_L}$ & 2745  & 2320   & 2810  \\
$m_{\tst_1}$ & 1636  & 1391  & 1760   \\
$m_{\tst_2}$ & 2258   & 1898  & 2120   \\
$m_{\tb_1}$ & 2255 & 1895  &   2130    \\
$m_{\tb_2}$ & 2730  & 2282  & 2400  \\
$m_{\te_L}$ &  2731  & 2294  & 2680   \\
$m_{{\tilde \tau}_1}$ & 2714  & 2255  & 2270   \\
$m_{{\tilde\chi_1}^{\pm}}$ &114 &113  & 255  \\
$m_{{\tilde\chi_2}^{\pm}}$ & 255 & 251 &434  \\
$\mlspfour$ & 255 &252  & 434  \\
$\mlspthree$ & 152 & 149 & 277  \\ 
$\mlsptwo$ & 136 &134 &  267 \\ 
$\mlspone$ & 82 &81 &  196  \\ 
$m_A$ & 2704  & 2212  &  1720  \\
$m_{H^+}$ & 2706  & 2214  &  1720\\
$m_h$ & 118 & 118 &119 \\
$\Omega_{{\tilde \chi}_1^0}h^2$ & 0.128 & 0.120 & 0.092 \\
$Br(b\to s\gamma)$ & $3.62\times 10^{-4}$ & $3.57\times 10^{-4}$ 
& $3.42\times 10^{-4}$ \\
$Br(B_s\to \mu^+\mu^-)$ & $3.12\times 10^{-9}$ & $3.11\times 10^{-9}$ 
& $3.02\times 10^{-9}$ \\
$\Delta a_\mu    $ & $4.60 \times  10^{-11}$ & $1.14 \times  10^{-10}$ 
& $2.56 \times  10^{-10}$ \\
\hline
\end{tabular}
\end{center}
\caption{\em mSUGRA Benchmark points A, B and C (masses are in GeVs). The 
first five parameters define the model, while the rest are predictions.}
\label{tabmsugra}
\end{table}

We also study the collider signatures for mSUGRA scenario at points with the
same (or very similar) gluino mass and $\tan\beta$ 
corresponding to the each of points A,B,C. 
These have been denoted by mSUGRA-A, mSUGRA-B and mSUGRA-C.
We must mention that the requirements of obeying  
the stringent WMAP data as well as the lower bound of the 
lighter chargino mass did not allow us to choose exactly identical  
values of the masses of the gluino in each case of the mSUGRA points.
This is particularly true for mSUGRA-A and mSUGRA-B that fall in the 
HB/FP zone.
The high scale parameters as well as the low scale soft masses for 
these points are listed in Table 2, all of them being consistent with the 
constraint from WMAP. $sign(\mu)$ is taken to be positive and the trilinear 
coupling $A_0$ is taken to be zero, as mentioned earlier. The corresponding 
low-energy spectra have also been generated via {\tt SuSpect} v2.34
using two-loop RGEs. 
Full one-loop and the dominant two-loop 
corrections to the Higgs masses are incorporated. We have used the strong 
coupling ${\alpha_3 (M_{Z})}^{\overline{MS}}= 0.1172$ 
for this calculation, adopting the default option in {\tt SuSpect}. 
We have assumed the top quark mass to be 172.7 GeV throughout the analysis, 
and no tachyonic sfermion mode has been allowed at any scale.
We now comment on 
the differences in spectra between the NUSM and mSUGRA 
benchmark points. For reasons that have been already stated, the high scale 
scalar mass parameters need to be chosen differently in the two cases. 
Consequently, the value of $\mu$ in NUSM is
larger than that in mSUGRA, simply because the mSUGRA benchmark points are 
within or very close to the HB/FP zones. We must note that there is no HB/FP 
like effect in NUSM that would reduce $\mu$. As a result, the chargino and 
neutralino masses in mSUGRA benchmark points are smaller than their 
counterparts in NUSM. 

Finally, in regard to the mass of gluino it is 
important to clarify the role of 
radiative corrections in the benchmark points of the two scenarios 
namely mSUGRA-$i$ and NUSM's point $i$, where $i\equiv$~A,B,C. 
Radiative corrections comprising of 
gluon-gluino and quark-squark loops may be 
estimated as in Eq.\ref{gluinomasseqn}\cite{Pierce:1996zz}.   
\begin{eqnarray}
m_{\tilde{g}} &=& m_{3}(Q^2) + {3\alpha_s\over 4\pi}
m_3\left(5-3\ln\left({{m_3^2\over Q^2}}\right)\right) \nonumber \\
&-& \sum_{q=u,..,t}
{\alpha_s\over4\pi}m_3\,{\rm Re}\left[
 \hat{B}_1(m_3^2,m_q^2,m^2_{\tilde{q}_1}) 
+ \hat{B}_1(m_3^2,m_q^2,m^2_{\tilde{q}_2})\right] \nonumber \\
&+&  \sum_{q=t,b}{\alpha_s\over4\pi}m_q\sin(2\theta_q)\,{\rm Re}\left
[B_0(m_3^2,m_q^2,m^2_{\tilde{q}_1})- B_0(m_3^2,m_q^2,m^2_{\tilde{q}_2})
\right]. 
\label{gluinomasseqn}
\end{eqnarray}
The Passarino-Veltman functions $B_0$, ${\hat B}_1$ and 
further useful details 
may be seen in Ref.\cite{Pierce:1996zz}.
The choice of the scale $Q$ is not unambiguous and, in general, 
   is defined by an appropriate mass scale in the theory. 
   In {\tt SuSpect}, for example, this is set equal to the geometric
   average of the values of the two stop squark masses. 
As we can see from Table~\ref{gluinomasstable},  this average 
varies widely between mSUGRA-$i$ and the corresponding NUSM's benchmark 
point $i$. As a result 
the running mass $m_3(Q^2)$ for mSUGRA-$i$ is smaller where $Q$ is higher 
compared to point $i$ of NUSM, where the corresponding scale 
is smaller because the 
masses of the third generation of squarks in NUSM are quite smaller
\footnote{$m_3(Q^2)$ increases with a decreasing $Q$: see for example  
Ref.\cite{IbanezNPB1984}.}. 
\begin{table}[!ht]
\begin{center}
\begin{tabular}{|c||c|c|c|c|c|c|}
\hline
Points & $m_0$ & $m_{1/2}$ & Q  & $m_3(Q^2)$ & $m_{\tilde g}$ & 
Radiative Correction \\ 
\hline
 & GeV & GeV & GeV & GeV & GeV &  \\ 
\hline
A & 2050 & 270 & 357 & 633 & 709 & 12\% \\ 
\hline
mSUGRA-A & 2740 & 253 & 1917 & 548 & 740 & 35\% \\ 
\hline
B & 2000 & 255 & 329 & 602 & 674 & 12\% \\ 
\hline
mSUGRA-B & 2300 & 252 & 1619 & 551 & 725 & 32\% \\ 
\hline
C & 1250 & 540 & 903 & 1197 & 1278 & 7\% \\ 
\hline
mSUGRA-C & 2680 & 490 & 1921 & 1051 & 1273 & 21\% \\
\hline
\end{tabular}
\end{center}
\caption{\em Running mass, radiative correction in percentage and pole mass of gluino in mSUGRA and NUSM benchmark points. The scale $Q$ refers to geometric mean of 
stop squark mass values.}
\label{gluinomasstable}
\end{table}
In general, for the given benchmark points under consideration,  
a point mSUGRA-$i$ has a smaller running mass $m_3(Q^2)$ 
but has a much larger contribution from radiative corrections 
({\em vide} Eq.\ref{gluinomasseqn}) compared to the 
corresponding point $i$ of NUSM. 
We note that the radiative correction amounts that arise from gluon-gluino  
and quark-squark loops are quite different in the  
two scenarios. With a heavier average squark mass, a benchmark point 
mSUGRA-$i$ has 
a much smaller contribution from quark-squark loops compared to that 
of NUSM point $i$. On the other hand, the logarithmic term in 
Eq.\ref{gluinomasseqn} is such that for mSUGRA-$i$ the term is negative 
because of the fact that $m_3^2(Q^2)<Q^2$ owing to a heavier average SUSY mass 
scale.  
This leads to a large contribution from the second term of the same 
equation for mSUGRA-$i$. This, 
however is not true for NUSM where one has $m_3^2(Q^2)>Q^2$ owing 
to a lighter average stop mass or a lighter SUSY mass scale in general.      
NUSM points also have significant amount of quark-squark contributions 
for the same reason.
The final effect is such that
smaller values of $m_3(Q^2)$ are overrun  
by radiative corrections in mSUGRA-A and mSUGRA-B leading to larger values of 
the pole masses $m_{\tilde g}$ in comparison to 
the values of $m_{\tilde g}$ for NUSM benchmark points A and B. 
The point mSUGRA-C and NUSM point C are quite competing in the above effects 
because of a larger associated $m_3$ and the values of $m_{\tilde g}$ are thus 
close to each other.  

\section {Collider Signatures}
\label{collidersection}

\subsection {The general strategy}

The collider signatures, and hence the optimal search strategies, of the NUSM
would naturally depend on the particular point in the parameter space that
nature may have chosen.  Rather than attempting a general, and hence
non-optimal, analysis, we choose to illustrate the various features, 
concentrating largely
on the three representative points identified in the preceding
section. To start with, we summarize, in brief, the generic 
simulation procedure that has been adopted here. The spectrum generated by
{\tt SuSpect} v2.34 as described earlier is fed into the event generator {\tt
Pythia} 6.4.16 \cite {Pythia} through a standard {\tt SLHA} \cite {sLHA} 
interface for the simulation of
$pp$ collisions with a centre-of-mass energy of 14 TeV.

We have used the {\tt CTEQ5L} \cite{CTEQ}
parton distribution functions, 
the QCD renormalization and factorization scales 
both being
set at the subprocess centre-of-mass energy $\sqrt{\hat{s}}$.
All possible SUSY processes and decay chains consistent 
with conserved $R$-parity have been kept open. We have kept 
initial and final state radiations (ISR/FSR) on. The effect of multiple 
interactions has been neglected though. We, however, take 
hadronization into account using the fragmentation functions 
built into {\tt Pythia}.

In Table \ref{tab:prodn}, we list the total supersymmetric particle production
cross-sections for each of the benchmark points. Also listed are the
individual cross sections for some of the important processes, 
namely, 
$\tilde g \tilde g$, $\tilde t_{1(2)} \tilde t^*_{1(2)}$ and $\tilde b_{1(2)} \tilde
b^*_{1(2)}$ 
and processes with at least one chargino or neutralino denoted by ``${\tilde \chi}_i^0/{\tilde \chi}_{1,2}^\pm$''.  
We note 
that, for points A and B, dominant production accrues from
stop pairs, while for point C, no production process dominates 
overwhelmingly. The
other important processes include associated stop and sbottom production
along with gluinos as well as charginos and neutralinos. That the total cross-section is
much smaller for point C, compared to the other two, is easy to understand as
the spectrum is relatively heavier in this case. 
It should be noted that while the mSUGRA and the NUSM benchmark points
    are quite similar as far as the gluino-pair production or the total 
    supersymmetric particle production cross sections are concerned, 
they differ 
    markedly in the dominant production modes. For the 
    mSUGRA points, it is the lighter neutralinos and charginos that 
    dominate overwhelmingly,
    whereas for the NUSM points, this r\^ole is usurped by stop-pairs 
    and sbottom pairs.   
\begin{table}[!ht]
\begin{center}
\begin{tabular}{|c|c|c|c|c|c|c|c|c|}
\hline
\multicolumn{4}{|c|}{mSUGRA} & \multicolumn{5}{|c|}{NUSM} \\
\hline
Point & Total & ${\tilde \chi}_i^0/{\tilde \chi}_{1,2}^\pm$ & {$\tilde g \tilde g$} & Point & Total & {$\tilde t_1 {\tilde t_1}^* + \tilde t_2 \tilde t^*_2$} & {$\tilde b_1 {\tilde b_1}^* + \tilde b_2 \tilde b^*_2$} 
& {$\tilde g \tilde g$}\\
\hline 
{\bf mSUGRA-A} & 11.86 & 10.67 & 1.18 & {\bf A} & 12.42 & 6.77 & 1.73 & 1.28 \\ 
\hline
{\bf mSUGRA-B} & 12.49 & 11.18 & 1.25 & {\bf B} & 19.92 & 11.73 & 2.79 & 1.78 \\ 
\hline
{\bf mSUGRA-C} & 0.62 & 0.59 & 0.02 & {\bf C} & 1.23 & 0.09 & 0.02 & 0.07 \\ 
\hline
\end{tabular}
\end{center}
\caption{\em Total supersymmetric particle production cross-sections 
(in pb) as well as the leading contributions for each of the NUSM and mSUGRA
benchmark points.}
\label{tab:prodn}
\end{table}

\begin{table}[!ht]
\begin{center}\
\begin{tabular}{|c|c|c|c|}
       \hline
Decay modes & A &B&C\\
(squark/gluino) & &&\\
\hline
\hline
 $\widetilde g \rightarrow \widetilde b_1 b$&31.0  &33.0&28.0\\
\hline
 $\widetilde g \rightarrow \widetilde b_2 b$&26.0  &26.0&20.0\\
\hline
$\widetilde g \rightarrow \widetilde t_1 t$ &22.0  &21.0&29.0\\
\hline
$\widetilde g \rightarrow \widetilde t_2 t$ &21.0  &20.0&23.0\\
\hline
\hline
$\widetilde b_1 \rightarrow \lspone b$ &8.0  &13.0&7.0\\
\hline
$\widetilde b_1 \rightarrow \lsptwo b$ &42.0  &52.0&24.0\\
\hline
$\widetilde b_1 \rightarrow \chonem t$ &11.0  &0.0&40.0\\
\hline
$\widetilde b_1 \rightarrow \lstop W^{-}$ &37.0  &33.0&3.0\\
\hline
\hline
$\lstop \rightarrow \chonep b$ &100.0  &100.0&33.0\\
\hline
$\lstop \rightarrow \ch2m b$ &0&0&21.0\\
\hline
$\lstop \rightarrow \lspone t$ &0.0  &0.0&23.0\\
\hline
$\lstop \rightarrow \lsptwo t$ &0.0  &0.0&13.0\\
\hline
\hline
$\lsptwo \rightarrow \lspone q \bar q$ &69.0&71.0&0.0\\
\hline
$\lsptwo \rightarrow \lspone l \bar l$ &10.0&10.0&0.0\\
\hline
$\lsptwo \rightarrow \lspone \nu \bar \nu$ &20.0&18.0&0.0\\
\hline
$\lsptwo \rightarrow \lspone h$ &0.0&0.0&90.0\\
\hline
\hline
$\chonep \rightarrow \lspone W^+$ &100.0&100.0&100.0\\
\hline
\end{tabular}

\end{center}
   \caption{\it {The branching ratios($\%$)of the dominant decay modes of the gluinos, squarks 
and lighter electroweak gauginos for NUSM for the different benchmark points.}}
\label{tab:NUSM_br}
\end{table}

\begin{table}[!ht]
\begin{center}\
\begin{tabular}{|c|c|c|c|}
       \hline
Decay modes & mSUGRA-A &mSUGRA-B&mSUGRA-C\\
(squark/gluino) & &&\\

\hline
 $\widetilde g \rightarrow \chonem t b$&26.6  &27.4&33.0\\
\hline
 $\widetilde g \rightarrow \chargino2 t b$&19.0  &19.0&12.0\\
\hline
$\widetilde g \rightarrow \lspi t \bar t$ &22.0  &22.0&31.8\\
\hline
\hline
$\lsptwo \rightarrow \lspone q \bar q$ &33.0&35.4&54.4\\
\hline
$\lsptwo \rightarrow \chonem u d$ &32.4&32.0&12.0\\
\hline
$\lsptwo \rightarrow  \chonem l \nu_l$ &10.0&15.4&6.0\\
\hline
\hline
$\chonep \rightarrow \lspone u d$ &66.6&66.6&66.6\\
\hline
$\chonep \rightarrow \lspone l \nu_l$ &33.0&33.0&33.0\\
\hline
\end{tabular}
\end{center}
   \caption{\it{The branching ratios($\%$)of the dominant decay 
modes of the gluinos,
 lighter neutralino and chargino states   
for mSUGRA for the different benchmark points.}}
\label{tab:msugra_br}
\end{table}

Before we discuss the signals, it behoves us to briefly discuss 
the major decay modes (see Tables~\ref{tab:NUSM_br} and \ref{tab:msugra_br}), for the structure of the cascades 
would determine the final state configurations. 
Starting with the major produce, 
namely the stop, 
for each of points A and B, it 
has almost a 100\% decay branching fraction to 
$b$ and ${\tilde \chi^+_{1}}$.
The ${\tilde {\chi}^{\pm}_{1}}$ decays,
in turn, into $W^\pm$ and the LSP again with nearly a 100\% 
branching fraction. 
For point C, although $\tilde{t_1}\longrightarrow 
b\tilde \chi^+_{1}$ is the dominant decay mode, the stop also has sizable 
branching into $t\tilde \chi^0_{1}$, 
$t\tilde \chi^0_{2}$ and $b\tilde \chi^{\pm}_{2}$.
As for the sbottoms, they have sizable branching fractions into 
both the top--chargino and the bottom--neutralino modes. The former, though 
slightly suppressed on account of phase space considerations,  
is particularly interesting in that 
it leads to tops in final states.  
With the stop and sbottom being so light in
this scenario, it is obvious that the gluino decay branching fractions into
stop and sbottom (accompanied by a top or a bottom, as the case may be) are
significantly enhanced as compared to the typical mSUGRA case. In fact, 
these modes, all of comparable magnitudes, together 
turn out to be overwhelmingly 
dominant. This, obviously, results in an enhanced scope of having top 
and/or bottom quarks 
with a high multiplicity. This, in turn, makes it likely to have several 
leptons in the final state (typically from the top quark decays). This 
particular character of the spectrum, thus, raises hopes for 4$\ell$ 
(with $\ell = e, \mu$) final states as a viable signal of SUSY. We, 
nonetheless, do not limit ourselves to these alone, but consider each
of the following final states:

\begin{itemize}
\item Opposite sign dilepton ($OSD$) :
$(\ell^{\pm}\ell^{\prime \mp})+ (\geq 2)~ jets~ + \met$ , 
  
\item Same sign dilepton ($SSD$) : 
$(\ell^{\pm}\ell^{\prime \pm})+ (\geq 2)~jets~ + \met$ ,

\item Trilepton $(3\ell+jets)$: 
$3\ell~ + (\geq 2) ~jets~ + \met$ ,

\item Hadronically quiet trilepton\footnote{These get contributions 
from electroweak production of a chargino and a neutralino.}  $(3\ell)$:
$3\ell~ + \met$ , 

\item Inclusive 4-lepton ($4\ell+X$): $4\ell + X + \met $,     
\end{itemize}

\noindent
where $\ell$ stands for final state electrons and/or muons, $\met$ denotes 
missing transverse energy and $X$ denotes any associated jet(s).

Of the various final states listed above, only the hadronically quiet 
trileptons have their origin in electroweak processes such as $\lsptwo \chonem$ 
production. However, as can be seen from our event selection criteria set down
in section \ref{det_kine}, strong processes which do not give rise
to hard enough jets can also be responsible for such final states. The large
rate of $t \tilde t$ and $b \tilde b$ production in NUSM thus leads to
relatively higher rates for hadronically quiet trileptons. On the whole,
rates are never found to exceed a few percent of those with accompanying 
hard jets.

As is well known, in the LHC environment, even if the hard scattering process
were to lead to a purely non-hadronic final state, the actual observable 
final state would, nonetheless, still include typically a few jets, 
originating from underlying events, pile up effects and ISR/FSR. In view of
this, we define a hadronically quiet event to be one devoid entirely 
of any jet with ${E_{T}}^{jet} ~\geq ~100$ GeV. 
This avoids unnecessary removal of events accompanied by 
relatively soft jets.

\subsection {Detection and Kinematical Requirements}
\label{det_kine}
Before we mention the selection cuts, we would like to discuss the resolutions 
of the detectors, specifically those applicable to the ECAL, the 
HCAL and  the muon 
chamber that have been incorporated in our analysis \cite {Sanjoy}. 
This is particularly 
important for reconstructing missing-$E_T$, which is a key variable for 
discovering physics beyond the Standard Model.  

We assume that all  charged particles with $p_T>$ 0.5 GeV 
are detected\footnote 
{This threshold is specific to ATLAS, while for CMS, $p_T>$ 1 GeV is 
applicable. Our results, though, are largely insensitive to the 
exact figure.} as long as they emanate
within the pseudorapidity range $|\eta|<5$.  
For muons though, the applicable pseudorapidity range
is determined by the geometry of the 
muon chamber to be\footnote{Although it seems that 
muons in the range $2.5<|\eta < 5$ would 
leave their footprints in the tracker, we deliberately choose to 
be consistent with the above criteria. Once again, the inclusion of 
such muons would make little quantitative difference.}
$|\eta| < 2.5$.
All the particles thus detected constitute the ``physics objects'' 
that are reconstructed in a collider experiments, and are further 
classified as

\begin{itemize}
\item isolated leptons;
\item hadronic jets formed after identifying isolated leptons; 
\item unclustered energy comprised of calorimetric
         clusters with $p_T>0.5$ GeV 
(ATLAS) and $|\eta|<5$, that are not associated with any of the above types of 
high-$E_T$ objects  
\end{itemize}

Electrons and muons with $p_T>$ 10 GeV and $|\eta|<$ 2.5 may be
identified as isolated leptons. In order to be deemed isolated, the
lepton should be sufficiently separated from any other lepton in that
it must satisfy ${\Delta R}_{\ell\ell} \geq $ 0.2, where $\Delta R =
\sqrt {(\Delta \eta)^2 + (\Delta \phi)^2}$ is the separation in the
pseudorapidity--azimuthal angle plane.  Similarly, it must be far away
($\Delta R_{\ell j} \geq 0.4$) from all putative jets with $E_T >$ 20
GeV. And, finally, the total energy deposit from all hadronic activity
within a cone of $\Delta R \leq 0.2$ around the the lepton axis should
be $\leq$ 10 GeV.

Jets are formed with all the final state particles after removing the isolated
leptons from the list with {\tt PYCELL}, the
 inbuilt cluster routine in {\tt Pythia}. The detector is assumed to
stretch over the pseudorapidity range   $|\eta| \leq5$ and is segmented into 100 equal-sized
(in $\eta$-spread) strips. Similarly, the entire $2 \pi$ azimuthal
spread is again segmented into 64 equal-sized strips resulting in a
$100 \times 64$ grid of cells.
To register a signal, a minimum $E_T$ of $0.5
\gev$ needs to be deposited in a cell, while the minimum $E_T$ for a cell to
act as a jet initiator is assumed to be 2 GeV. All 
 objects within a cone
of $\Delta R$=0.4 around the jet initiator cell are considered for 
 jet
formation, and
for a conglomeration to be considered a jet, it must satisfy
$\sum_{\rm objects} E_{T} > 20 \gev$.

Now, as has been mentioned earlier, all the other final state
particles, which are not isolated leptons and are yet separated from
jets by $\Delta R \ge$ 0.4 are considered as unclustered objects. This
includes all electromagnetic objects (muons) with $0.5 \gev < E_T < 10
\gev$ and $|\eta|< 5 \, (2.5)$ as well as hadronic energy deposits
with $0.5 \gev < E_T < 20 \gev$ and $|\eta|< 5$.  Such unclustered
energy deposits need to be taken into account in order to properly
reconstruct any missing-$E_T$.

Any detector suffers from finite resolutions and collider 
detectors are no exception. To approximate the attendant experimental 
effects, we smear the energies (transverse momenta) 
with Gaussian functions. 
Nominally, the widths of the distributions 
have different contributions (accruing from different sources), each with 
a characteristic energy dependence and with these being added in quadrature.
To wit (all energies are measured in units of GeV),
\begin{itemize}
\item {electron/photon energy resolution:}
\begin{subequations}
\be
\frac{\sigma(E)}{E} = \frac{a}{\sqrt{E}} \oplus b \oplus
                      \frac{c}{E}
\ee
where
\be
(a, b, c) = \left\{ \begin{array}{lcl}
                       (0.030, 0.005, 0.2) & \hspace{1em} & |\eta| < 1.5 \\
                       (0.055, 0.005, 0.6) & \hspace{1em} & 1.5 < |\eta| < 5
		    \end{array}
            \right.
\ee
\end{subequations}
\item { muon $p_T$ resolution :}
\begin{subequations}
\be
\frac{\sigma(P_T)}{P_T} = \left\{ \begin{array}{lcl}
                       a  & \hspace{1em} & |\eta| < 1.5 \\
                       \displaystyle
                         a + b \log \frac{p_T}{100 \gev}  & & 1.5 < |\eta| <2.5
		    \end{array}
            \right.
\ee
with
\be
(a, b) = \left\{ \begin{array}{lcl}
                       (0.008, 0.037) & \hspace{1em} & |\eta| < 1.5 \\
                       (0.020, 0.050) & \hspace{1em} & 1.5 < |\eta| < 2.5
		    \end{array}
            \right.
\ee
\end{subequations}

\item {jet energy resolution :}

\be
\frac{\sigma(E_T)}{E_T} = \frac{a}{\sqrt{E_T}}
\ee
with 
$ a= 0.55 $ being the default value used in {\tt PYCELL}

\item {unclustered energy resolution :}

\be
\sigma(E_T)=\a\sqrt{\Sigma_{i}E^{(Unc.E)_i}_T}
\ee
where $\a\approx0.55$. One should keep in mind here 
that the $x$-- and $y$--components of 
$E^{Unc. E}_T$ need to be smeared independently (with 
identical widths). 

\end{itemize}

Once we have identified the `physics objects' as described above, we sum
vectorially the transverse components of all the momenta smeared thus 
to obtain the total visible transverse momentum.  Clearly, the 
missing transverse energy is nothing but the magnitude 
of the visible transverse momentum, namely
\be
\met =\sqrt{\left(\Sigma \,p_x \right)^2+\left(\Sigma p_y\right)^2}
\ee
where the sum goes over all the isolated leptons, the jets as well 
as the unclustered energy deposits. At this stage, we are in a 
position to impose the selection cuts, namely
\begin{itemize}
\item  Missing transverse energy $\met \geq 100 \gev$, 

\item $p_{T}^\ell ~\ge ~20$ GeV for all isolated leptons, 
 
\item $E_{T}^{\rm jet} ~\geq ~100$ GeV and $|{\eta}_{jet}| ~\le ~2.5$, 

\item For the hadronically quiet trilepton events, as
 also for 
inclusive 4$\ell$ events, we 
reject, in addition, any event with a same flavour opposite sign lepton pair
satisfying $|M_{Z}-M_{\ell^{+}\ell^{-}}| \leq 10$ GeV. Such events are 
characterised by the requirement of having no central jet with $E_T>100$ GeV.

\end {itemize}

We have generated the corresponding SM backgrounds (with identical
kinematical cuts) using {\tt Pythia}. The bulk of the contribution comes from
$t \bar t$ events. To take into account the next to leading 
order (NLO) and next to leading log resummed (NLL) corrections---not
included in {\tt Pythia}---we rescale the results by the appropriate 
$K$-factor\cite{ttbar} viz. 2.23. Exclusive diboson ($WW$, $WZ$, $ZZ$)  
production constitutes another potential background, but it is easy to 
see that except for the hadronically quite trilepton channel, these 
contributions are very sub-dominant. Furthermore, these 
are reduced drastically by the cuts imposed, especially 
by the one on the leptonic invariant mass. Inclusive, i.e. including 
(multi-)jets, gauge boson production is another very serious background,
but can be estimated with a high accuracy using {\sc Alpgen}~\cite{alpgen}.
The combination of a large missing $E_T$ along with the requirement of 
at least two high-$p_T$ leptons reduces even this to innocuous levels.

\subsection {Results}
    \label{sec:results}
The event rates in the various channels discussed in the preceding 
section would, of course, differ amongst themselves 
and also depend on the 
point in the parameter space, both on account of the 
differing production cross sections and branching fractions
as well as the kinematical restrictions imposed. 
In Table \ref{tab:event_rates}, we tabulate the event rates in 
different channels obtained for the points A, B and C of the NUSM scenario 
as well as those for the corresponding mSUGRA ones. Also shown 
are the respective SM backgrounds. 

\begin{table}[!h]
\begin{center}

\begin{tabular}{|c|c|c|c|c|c|}
\hline
Model Points & $\sigma_{OSD}$ & $\sigma_{SSD}$ &$\sigma_{3\ell+jets}$& 
$\sigma_{3\ell}$ & $\sigma_{4\ell}$ \\
\hline
\hline 
 {\bf A} & 103 & 24.0 & 14.9 & 3.1 & 3.1 \\
\hline
 {\bf mSUGRA-A} & 33.7 & 15.4 & 8.1 & 0.4 & 1.3 \\
\hline
 {\bf B} & 135 & 28.7 & 19.0 & 4.4 & 3.8  \\
\hline
 {\bf mSUGRA-B} & 38.9 & 16.9 & 9.1 & 0.4 & 1.5   \\
\hline
 {\bf C} & 23.9 & 7.3 & 2.9 & 0.1 & 0.3 \\
\hline
 {\bf mSUGRA-C} & 1.8 & 0.6 & 0.4 & 0.1 & 0.1   \\
\hline
\multicolumn{6}{|c|}{SM Backgrounds}
\\
\hline
{\bf $t\bar t$} & $1.10 \times 10^3$ & 18.1 & 2.7 & 5.3 & 0.0 \\
\hline
{\bf $ZZ$,$WZ$,$ZH$,$Z\gamma$} & 16.3 & 0.3 & 0.5 & 1.1 & 0.4 \\
\hline
{\bf Total Bkgd} & $1.12 \times 10^3$ & 18.4 & 3.2 & 6.4 & 0.4 \\
\hline
\hline
\end {tabular}\\
\end{center}
\caption{\em Event-rates (fb) after cuts for non-universal 
scalar mass points and corresponding mSUGRA points with same gluino mass.
{\tt CTEQ5L} parton distribution functions used with $\mu_{F}=\mu_{R}=~ \sqrt{\hat s}$.}
 \label{tab:event_rates}
\end{table}


For the NUSM benchmark points, the gluino decays dominantly into top-stop and 
bottom-sbottom (see table 4). The source of leptons in the final state can 
thus be both the stop and 
the sbottom which can lead to the top and chargino in the next stage of the
cascade. Of course, appropriate branching and combinatoric factors are to be 
used in each case.
For the mSUGRA benchmark points, on the other hand, the gluino (which is the 
lightest strongly interacting superparticle) decays primarily into the 
three-body channels such as $t \bar t \lspi$ and $t b  \charginoi$ 
(see table 5). It should be 
remembered, however, that gluino decays mediated by light squark flavours are 
not entirely negligible and in fact, can account for upto one-third of 
the decays. This causes an effective enhancement, in the NUSM cases, 
of decays into the intermediate states containg top/chargino. 

Another crucial difference is the splitting between the 
$\tilde\chi_{1}^{\pm}$ and $\tilde\chi_{1}^{0}$ states. With the 
splitting being large in the NUSM case, the $W^{\pm}$ from 
$\tilde\chi_{1}^{\pm}$ can be 
nearly on-shell (as opposed to an off-shell one in the mSUGRA case), thereby 
resulting in typically harder leptons. Together, these features are responsible 
for effectively reducing the rates for leptonic final states for mSUGRA in 
comparison with the NUSM benchmark points consistent with the dark matter 
constraints. Of course, the already mentioned difference in gluino masses 
has also a small role to play.  

Based on the above observations, the following features in the results are 
noted.  
\begin{itemize}

\item For each of $OSD$, $3\ell$ and $4\ell$ final states, 
the difference in the absolute
rates between the NUSM points 
and the corresponding mSUGRA ones  is remarkably large. As 
has been argued at the beginning of Section 3.1, this 
can be understood in terms of the relative lightness of the third-generation
squarks in the NUSM scenario. 
While it might be tempting to aver that 
this alone would serve to distinguish NUSM from mSUGRA scenarios, 
a little reflection shows that just the absolute rates are not 
enough for this purpose and a 
combination of observables would be required.

\item For the same sign dilepton 
final state (a manifestation of the Majorana nature of the gauginos), 
the signal to background ratio ($S/B$) exceeds unity for both 
points A and B.  
For point C, while $S/B < 1$, a discovery is possible with an integrated 
luminosity of only $10 \fb^{-1}$. This, though, is not surprising, for 
$SSD$ is well known for its efficacy in SUSY search. Note though that the
$SSD$ rate cannot really distinguish between NUSM and its mSUGRA counterpart.
While the case for point C may look promising, it is precisely the case 
where the rate (and, hence, statistical significance) is low.

\item While 
$S / B \gsim 1$ for the 4-lepton final state as well,
the smaller rates for this signal significantly reduces its potential as 
a discovery channel.
However, this 
could potentially serve
 as a very efficient discriminator between scenarios.
  For example, the $S / B$ ratio is close to 
unity even for point C. 

\item The situation for the $3 \ell + {\rm jets}$ is somewhat better
      than the $4 \ell$ one. 
  The rates are larger while maintaining the difference 
between NUSM and mSUGRA. Once again, for point C, the 
$S/B$ ratio is close to unity, while for mSUGRA-C, it is suppressed.

\item As for the $OSD$ and the hadronically-quiet $3 \ell$ 
       final states, generically, 
      $S/B < 1$. The former though boasts of the largest event rates. For
      points A and B, this signal begins to stand well over the background
      fluctuation for an integrated luminosity of 
      as little as $1 \fb^{-1}$ whereas
      $2 \fb^{-1}$ would allow a discovery claim.
      For the hadronically-quiet trilepton mode, the required luminosity 
      is $\sim 10 \fb^{-1}$ for points {A} and {B}. 
      Qualitatively, these two points are very similar to each other, 
      especially as far as the superpartner masses are concerned. 
      The main difference lies in the Higgs sector, which 
      has not been explored here.

\item For point {C} (and even more so for mSUGRA-C), the heaviness 
      of the spectrum translates to lower event rates.

\item It is obvious that, for an integrated luminosity of
      $30 \fb^{-1}$, many of these channels would allow us to register
      a $5 \sigma$ discovery claim\footnote{The required luminosity is 
       much
      smaller for some channels and parameter points.}. In Table 
      \ref{tab:5sig}, we summarise this information for each of the channels 
      and parameter points, both in the NUSM scenario as well as their 
      mSUGRA counterparts.

\begin{table}[!htb]
\begin{center}

\begin{tabular}{|c|c|c|c|c|c|}

\hline
Model Points & ${OSD}$ & ${SSD}$ &${3\ell+jets}$& 
${3\ell}$ & ${4\ell}$ \\
\hline
\hline 
 {\bf A} & $\surd$ & $\surd$ & $\surd$ & $\surd$ 
& $\surd$ \\
\hline
 {\bf mSUGRA-A} & $\surd$ & $\surd$ & $\surd$ & $\times$ & 
$\surd$ \\
\hline
 {\bf B} & $\surd$ & $\surd$ & $\surd$ & $\surd$ & 
$\surd$  \\
\hline
 {\bf mSUGRA-B} & $\surd$ & $\surd$ & $\surd$ & $\times$ & 
$\surd$   \\
\hline
 {\bf C} & $\surd$ & $\surd$ & $\surd$ & $\times$ & $\times$ \\
\hline
 {\bf mSUGRA-C} & $\times$ & $\times$ & $\times$ & $\times$ & $\times$   \\
\hline
\hline
\end {tabular}
\end{center}
\caption{\em $5 \sigma$ visibility of various signals for an
	integrated luminosity of $30 \fb^{-1}$. 
	A $\surd$ indicates a positive conclusion while a
	$\times$ indicates a negative one.}
\label{tab:5sig}
\end{table}

\end{itemize}

\begin{figure}[!h]
\begin{center}
\centerline{\psfig{file=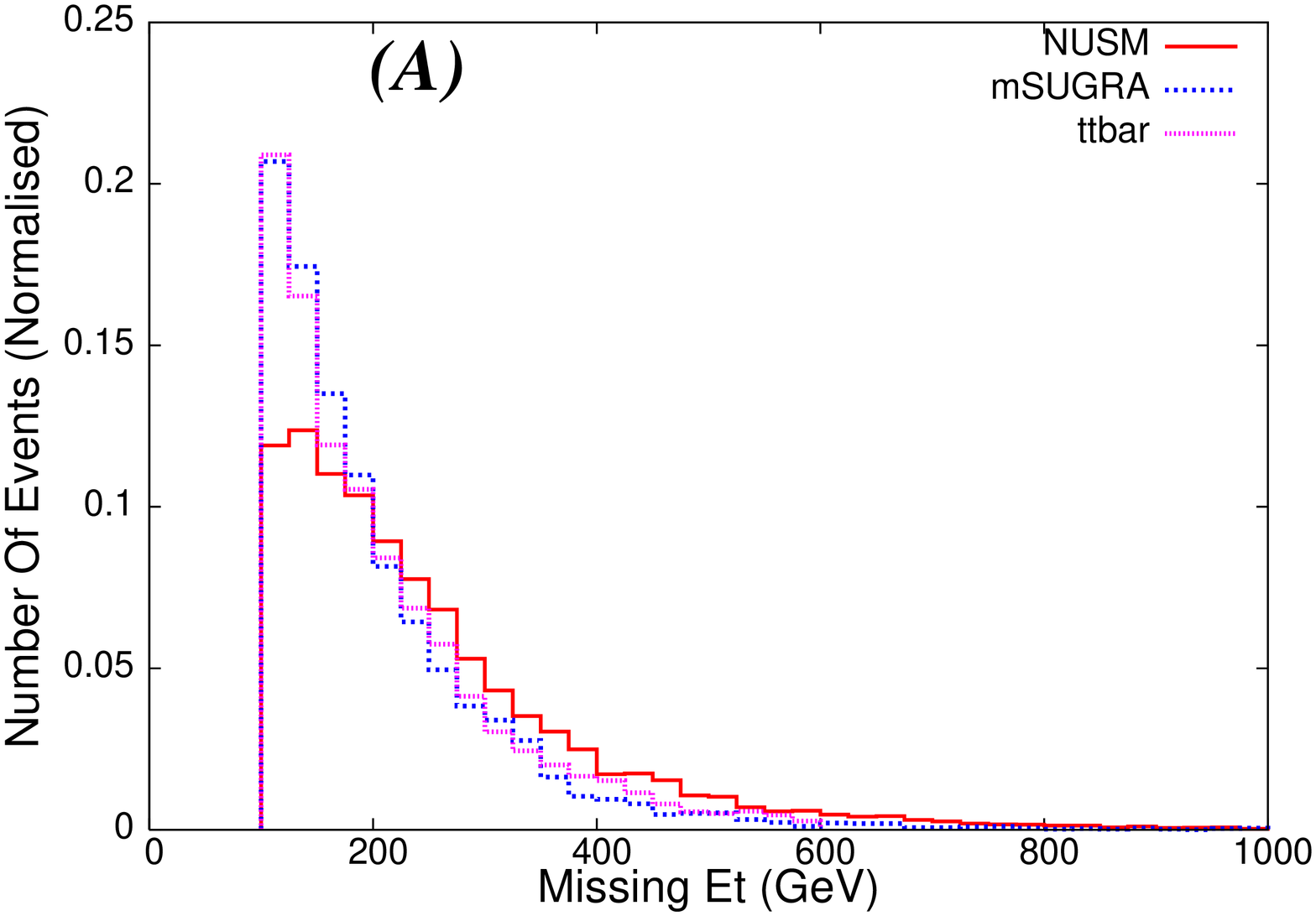,width=6. cm,height=6cm,angle=-90}
\hskip -15pt \psfig{file=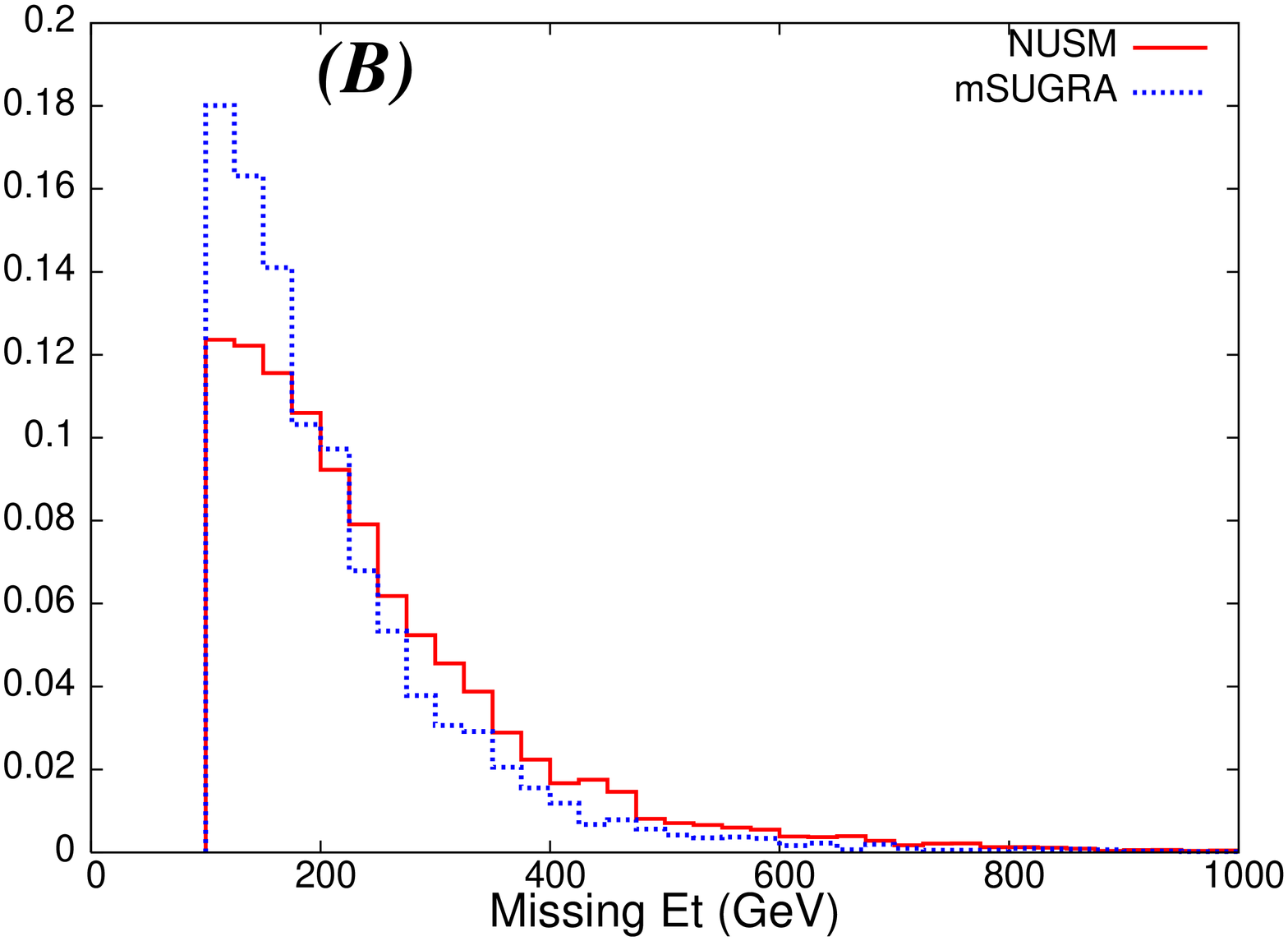,width=6. cm,height=6cm,angle=-90}
\hskip -15pt \psfig{file=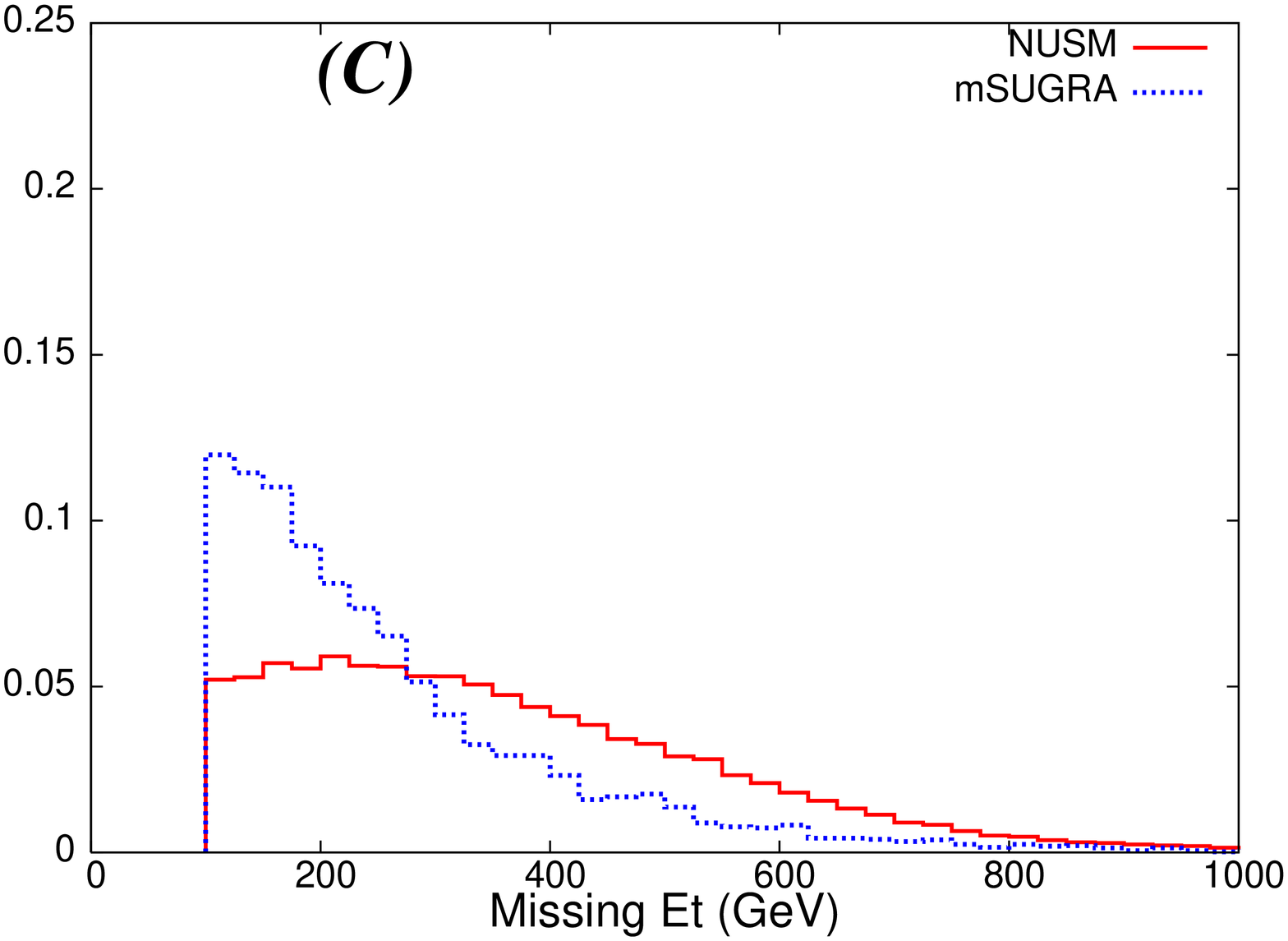,width=6. cm,height=6cm,angle=-90}}
\caption{\em Missing transverse energy distribution (normalized to unit area) 
for opposite sign dilepton ($OSD$) events. The eponymous panels refer to the 
respective representative points in the parameter space. Also shown are the 
analogous distributions for the corresponding mSUGRA points. The first 
panel also shows the distribution accruing 
from the overwhelmingly dominant SM background, 
namely $t \bar t$ production.} 
\label{mte}
\end{center}
\end{figure}

We now discuss the profile of the dominant
(though not necessarily the most background-free) 
signal mode, namely events with opposite
sign dileptons. The quest is to see if quantitative features in the same could
be used to either accentuate the discovery potential or as discriminators
between models and/or parameter points.  In Fig.\ref{mte}, we display the
normalized (to unity) distributions of missing transverse energy, a most
crucial aspect of supersymmetry signals. As a comparison of the first two
panels shows, the $\met$ distributions for parameter points A and B look very
similar, which is but a consequence of the aforementioned similarity 
between the corresponding spectra. Furthermore, all of them are discernibly 
different from those for the corresponding mSUGRA points\footnote {Note that 
we are concerned here about the {\em shape} of the curve, not the absolute 
magnitude, which, of course, are different ({\em vide} Table
\protect\ref{tab:event_rates}.)}. That the latter are softer can be understood
by realizing that the main production channel for the mSUGRA spectrum is $pp
\to \tilde g \tilde g$ and that, unlike in the NUSM case, the gluino undergoes
a three body decay, resulting in relatively less momentum imparted to the
LSP. Note also that the the dominant ($t \bar t$) background---as displayed
in the first panel---is almost as soft
as the mSUGRA signals, and thus a hardening of the $\met$ cut would have
considerably improved the $S/B$ ratio for the NUSM cases, while worsening it
for the mSUGRA ones. And, finally, 
for point {C}, the difference is
even more stark. However, with the size of the signal being small in this
case, it is not immediately apparent whether this could be used to any
advantage.

\begin{figure}[!ht]
\begin{center}
\centerline{\psfig{file=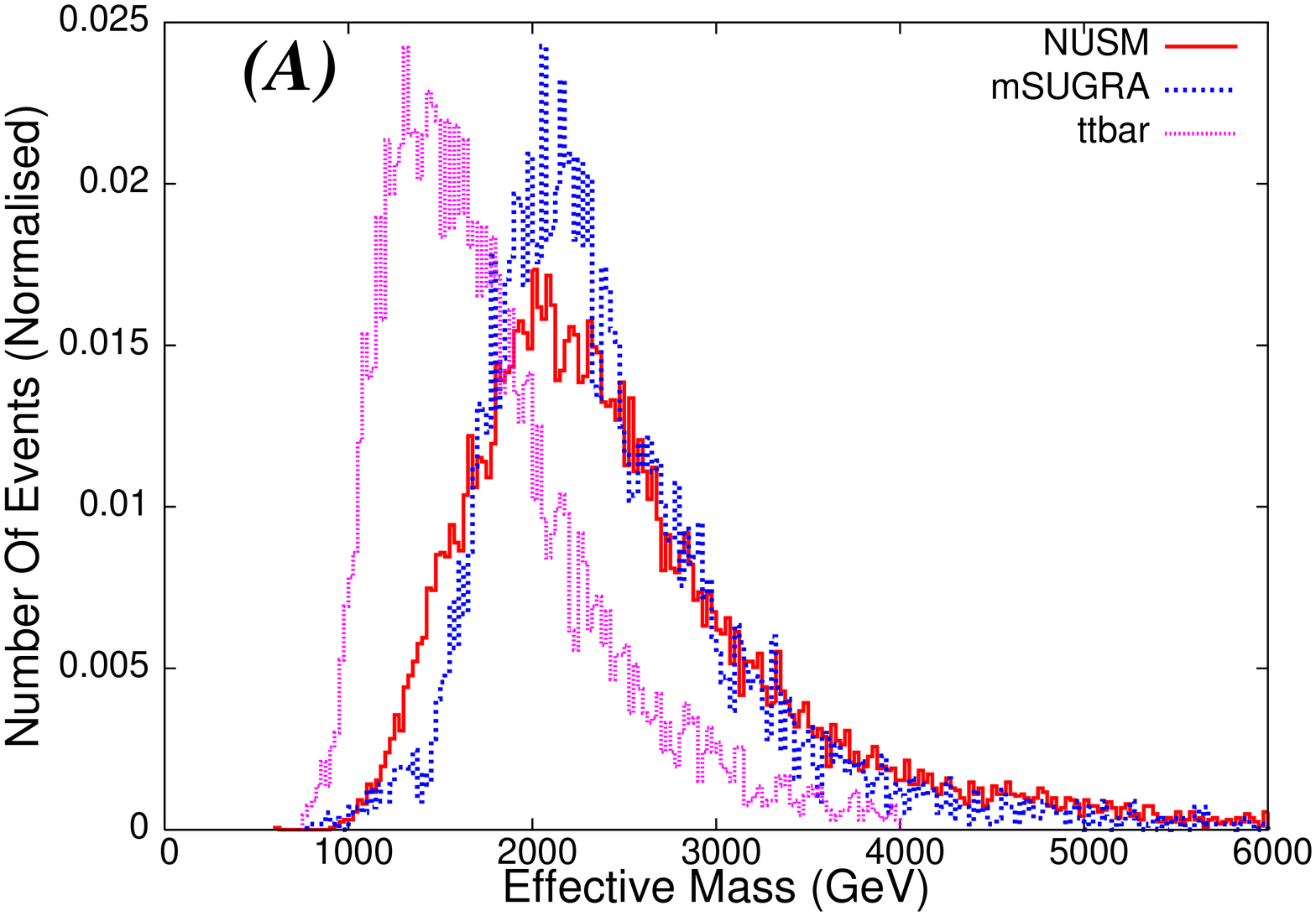,width=6 cm,height=6cm,angle=-90}
\hskip 0pt \psfig{file=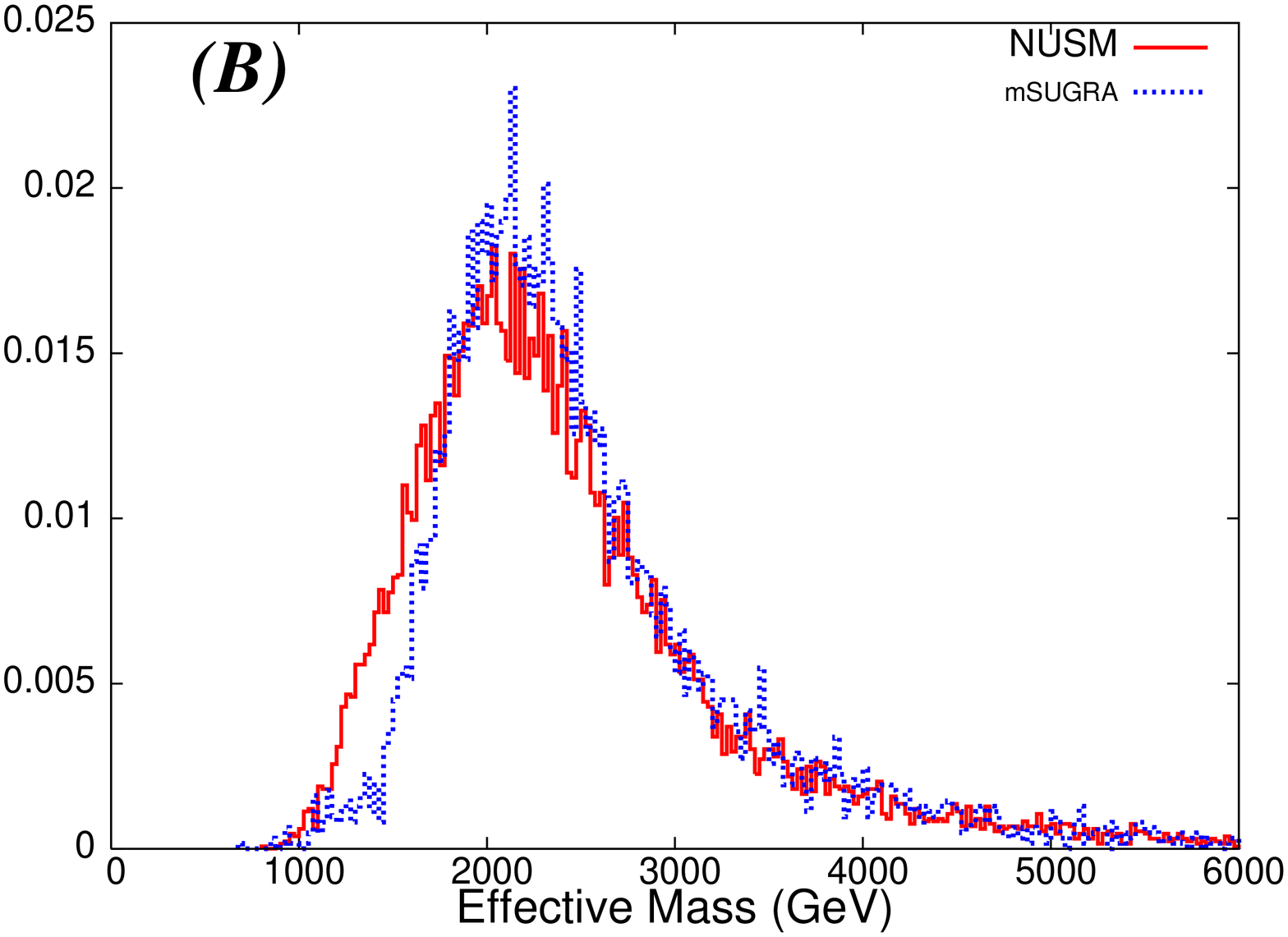,width=6 cm,height=6cm,angle=-90}
\hskip 0pt \psfig{file=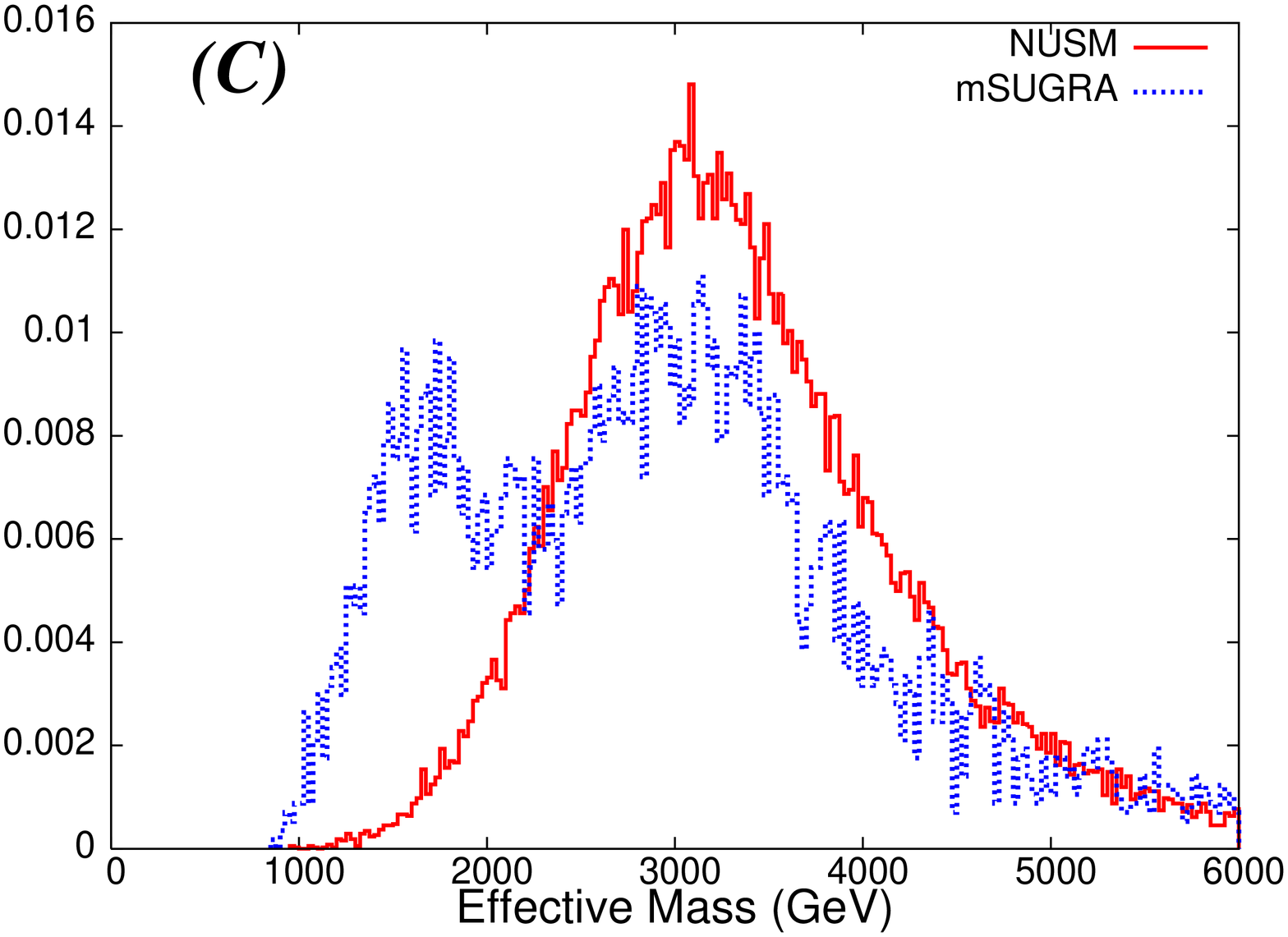,width=6 cm,height=6cm,angle=-90}}
\vskip 10pt
\caption{\em As in Fig.\protect\ref{mte}, but for the effective mass 
instead.}
   \label{fig:m_eff}
\end{center}
\end{figure}

Another kinematical variable often used advantageously in searches 
for new physics involving $\met$ is the ``effective mass'' defined 
to be the scalar sum of the transverse momenta 
of the isolated leptons and jets and the missing transverse energy, viz.
\bea
m_{\rm eff} \equiv \sum (p_T)_{{\rm iso.}\ell}+\sum (p_T)_{jets} + \met \ .
\eea
In Fig.\ref{fig:m_eff}, we display the corresponding distributions, again for
both the NUSM points and their mSUGRA counterparts.  As the first panel shows,
as far as point A is concerned, there is little to choose between this
distribution and that for the corresponding mSUGRA point. Similar is the case
for point B (second panel). Note, furthermore, that the peak in either case is
at a fairly large value of $m_{\rm eff}$.  While this, at first sight, might
seem contradictory to the oft-repeated claim that this distribution should
peak roughly at twice the mass of the dominant particle being produced, the
reason for this discrepancy is easy to appreciate. First and foremost, with the
strong demands made on the transverse momenta of the two leading jets, the
contribution from stop-pair production reduces drastically. This is
understandable since the relatively small difference between the stop mass and
those of the lighter chargino implies that the $b$ from stop decay tends to be
softer. With the stop-pair contribution thus being effectively 
decimated\footnote{It might seem paradoxical that we are altogether 
sacrificing the signal from the light stop, a cornerstone of this 
scenario. However, including the stop contribution would require softening the
$p_T$ requirements, a process fraught with danger in the context of the LHC. 
In the absence of a full-scale simulation including multiple scattering and
underlying events, we deliberately desist from this.}, this also offers
hints as to why the NUSM and mSUGRA distributions look so similar. And, with
the gluinos themselves being produced with a considerable transverse momentum, 
it is easy to understand why the distribution peaks at a high value
of $m_{\rm eff}$.

For point {C} though, a remarkable difference with the mSUGRA 
counterpart is immediately apparent. This can be traced to the fact 
that the charginos are considerably softer for mSUGRA-C as compared 
to the point C. This allows for the second, subdominant, hump slightly above 
twice the mass of the chargino. While this could, in principle, be 
used to discriminate between the two scenarios, unfortunately the smaller 
rates tend to make the task a difficult one. And, finally, note that 
the distribution for the (dominant) $t \bar t$ background is considerably
softer than those for any of the six supersymmetric parameter points 
discussed here. Thus, imposing a requirement such as $m_{\rm eff} \gsim 2
\tev$ would have significantly improved the $S/B$ ratio for the 
$OSD$ signal. This, though, would have eliminated the secondary hump 
for the mSUGRA-C case.

\begin{figure}[!h]
\begin{center}
\centerline{\epsfig{file=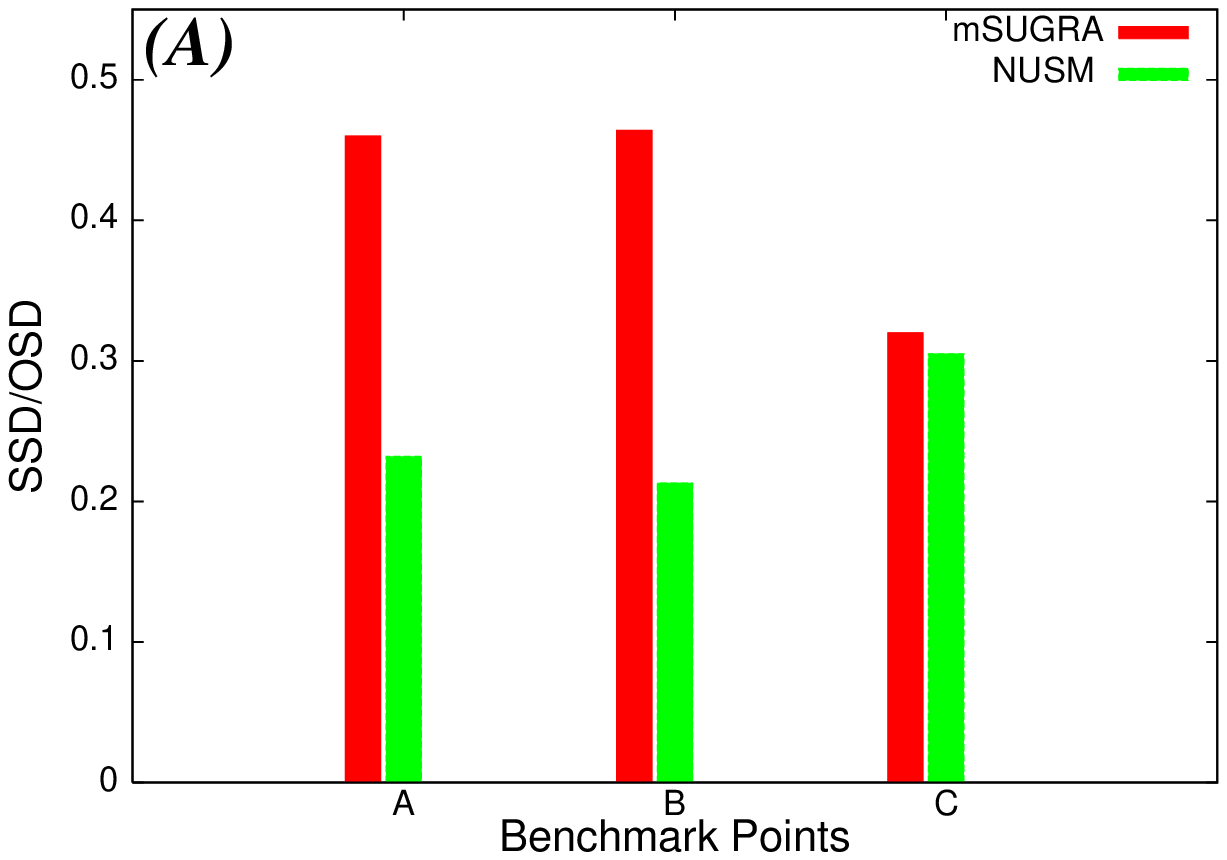,width=6.5 cm,height=5.50cm,angle=-0}
\hskip 20pt \epsfig{file=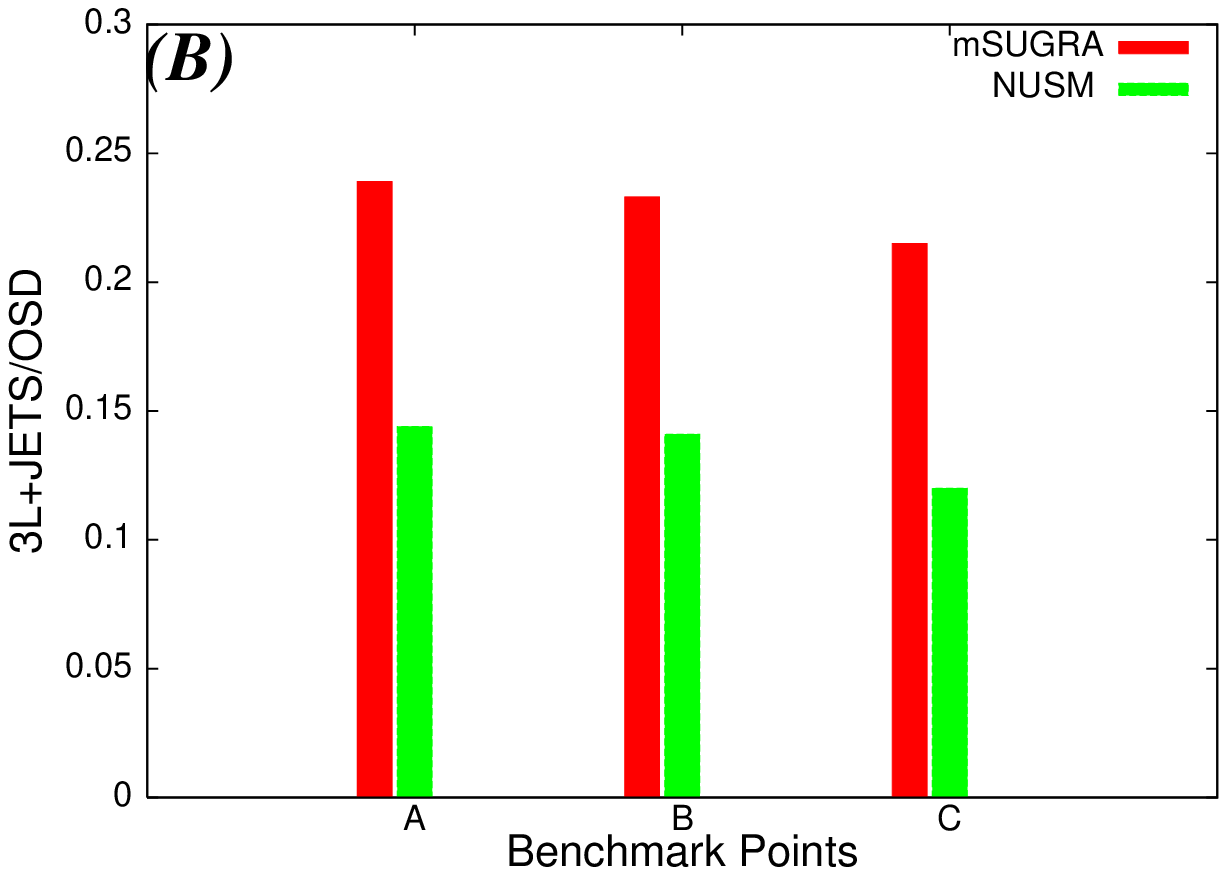,width=6.5cm,height=5.50cm,angle=-0}}
\vskip 10pt
\centerline{\epsfig{file=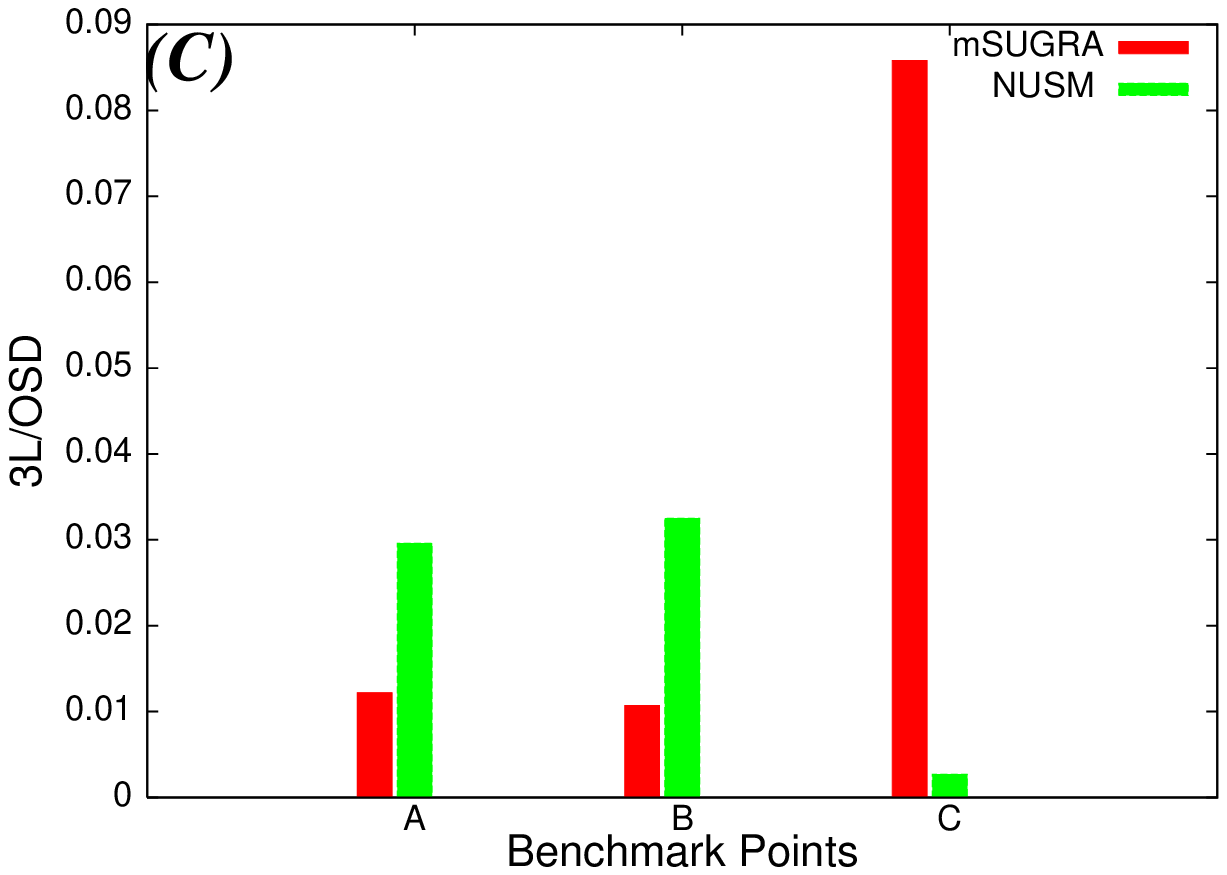,width=6.5 cm,height=5.50cm,angle=-0}
\hskip 20pt \epsfig{file=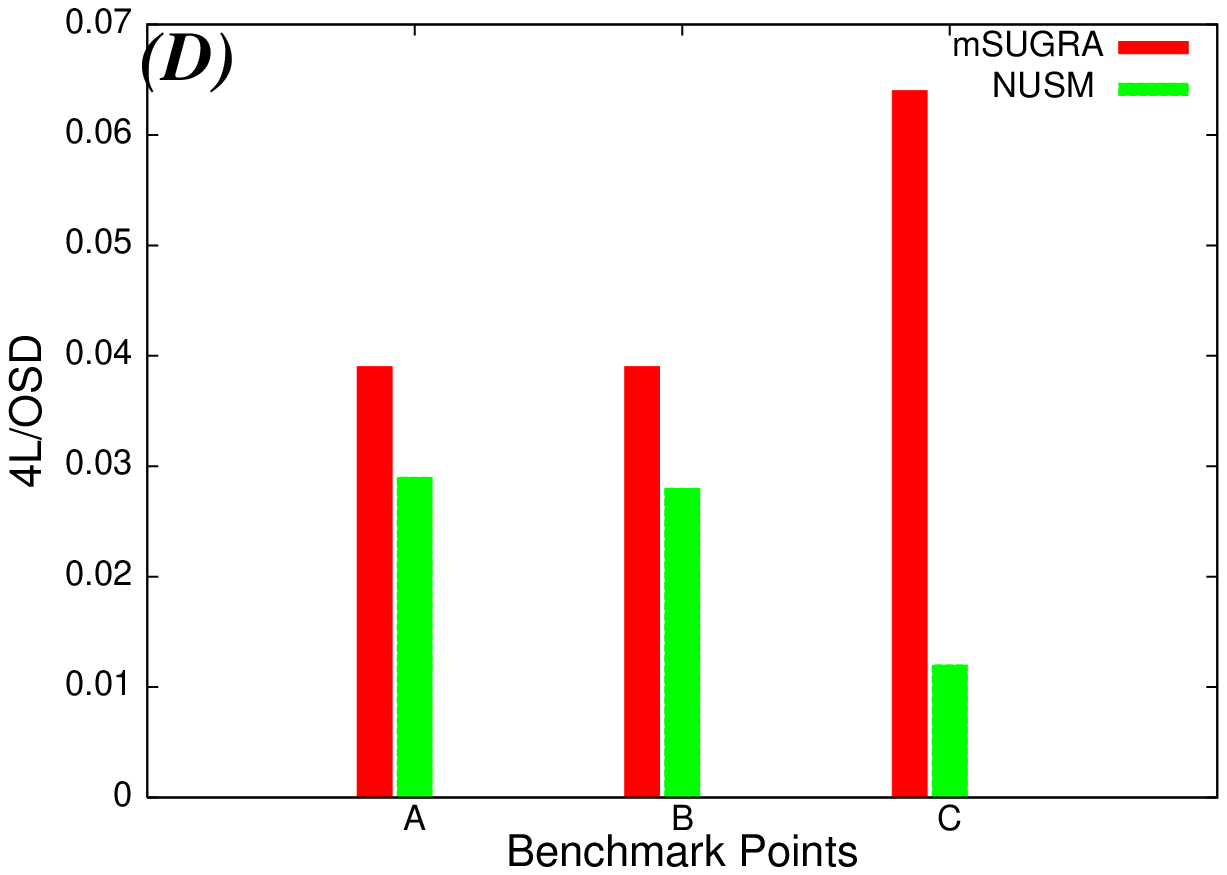,width=6.5cm,height=5.50cm,angle=-0}}
\end{center}
\caption{\em Event ratios with respect to opposite-sign dilepton($OSD$) events 
for NUSM and mSUGRA cases at the points A,B,C mentioned in the text.
{\it Colour Code: {\tt Red:} mSUGRA, {\tt Green:} NUSM scenario}. 
} 
\label{fig:ratio}
\end{figure}

Having discussed the prospects of refining and/or using the kinematical
variables in the $OSD$ sample towards discriminating between scenarios, we now
consider a set of observables, namely the ratios of events seen in
various channels.  
As is well known, there is a great advantage to the use of such
variables in that it almost entirely eliminates some systematic uncertainties
such as that in the luminosity and drastically reduces others such as those
corresponding to the choice of the parton distributions, the choice of
renormalization and fragmentation scales etc. In Fig.\ref{fig:ratio}, we
present the ratio of the subordinate channels with the dominant ($OSD$) 
channel for each of the parameter points. 

At this stage, we can easily formulate the means of discriminating 
between a NUSM point and the corresponding mSUGRA one, namely
\begin{itemize}
\item[\Large $\star$] For 
             parameter points (such as {A} and {B})
	     with a  relatively smaller  $\mhalf$ but large $m_0$ 
	     (i.e., when the 
             gluino is considerably heavier than the stop/sbottom but 
             sufficiently lighter than the first two generation squarks and 
             sleptons), the NUSM scenario would typically result in 
	     a smaller proportion of same sign dilepton events, as is 
             clear from Fig.\ref{fig:ratio}{\em (a)}, when compared 
	     to the $OSD$ rates. 	     
	     This can be attributed to the fact that the $OSD$-rate
             increases significantly for the gluinos decaying through stop-top
             (with, consequently, $OSD$ being 
	     possible even from the decay of a single gluino, irrespective
	     of how the other one decays), 
	     whereas the $SSD$ relies on the good old fact of the 
             gluino being a Majorana spinor, with only
	     a slight increment to the 
             leptonic branching fraction due to the decays through third 
             generation.

\item[\Large $\star$]
The mSUGRA-A and mSUGRA-B points have sufficiently small values of 
$\mu$ and these points indeed fall in the HB/FP region. This implies 
that there is more of Higgsino in
the lighter chargino (${\tilde \chi}_1^\pm$) 
and the second lightest neutralino (${\tilde \chi}_2^0$). Therefore, 
the leptonic signals are weakened compared to what we predict in
NUSM.

\item[\Large $\star$] Again, for 
             points with a small $\mhalf$ but large $m_0$
	     (such as {A} and {B}),
	     the rates (absolute and relative) for the 
	     hadronically quiet trilepton mode are markedly higher 
	     for the NUSM case. This can be attributed to the 
	     aforementioned feature of the NUSM spectrum
	     which renders it easier to have large-$p_T$ 
	     isolated leptons. The turnaround for point {C} vs 
	     mSUGRA-C is a consequence of the lightness of the 
	     charginos in the latter allowing a large contribution 
	     through the production of the charginos and the second-lightest
	     neutralino.

\item[\Large $\star$] Overall, it is self-evident that a combination
	     of these ratios would serve to easily distinguish between the 
	     two scenarios.

\item[\Large $\star$] And, finally, the relative lightness of the stop 
	     and the sbottom (and the consequent fact of the gluinos decaying 
	     through these), renders the NUSM signal $b$-rich. 
	     Invoking $b$-tagging (which we had
	      not done in the 
             results presented so far) would thus present us
	     with a very useful discriminator. 
	     With this in view, we 
             perform a study in $OSD$ channel associated with two or more 
             partonic $b$-jets (${\ge}2b+OSD$). 
             We assume a $b$-tagging efficiency\cite{b-tagging-ref} 
             of $\epsilon_b = 0.5$
	      for $p_T >$ 40 GeV and $|\eta| <$ 2.5. 
             The $OSD$ event selection criteria remain the same. As expected, 
             we see a clear distinction between the NUSM and the corresponding 
             mSUGRA ones in the absolute event rates as shown in Table 9.
	    With the NUSM sample being particularly rich in $b$'s,
	     the suppression in rates as compared to those in 
	     Fig.\ref{tab:event_rates} is understandably less severe 
	     than $\epsilon_b^2$. This, of course, 
	     does not apply as well to point {C} and far less 
	     to the MSSM cases, resulting in a large suppression 
	     of the latter.
	   
\begin{table}[!htb]
\begin{center}

\begin{tabular}{|c|c|c|c|c|c|c|c|}

\hline
Model Points &{\bf A}  &{\bf mSUGRA-A}  &{\bf B}& 
{\bf mSUGRA-B} & {\bf C} & {\bf mSUGRA-C} & $t\bar{t}$ \\
\hline
\hline 
 ${\ge}2b+OSD$ & 36.6  & 6.3  & 46.4 & 10.2 & 6.7 & 0.5 & 148.7 \\
\hline
\hline
\end {tabular}
\end{center}
\caption{{\em Event rates (fb) at different benchmark points and for the 
$t\bar t$ background for a final state ${\ge}2b+OSD$.}}
\label{tab:b-OSD}
\end{table}
	      
\end{itemize}

\section {Summary and Conclusions}
\label{conclusionsection}
We have studied a case of nonuniversal scalar masses, wherein the
first two families of squarks as well as sleptons of all generations 
are much heavier than 
the third family of squarks and the Higgs scalars. 
The universality of gaugino masses has been adhered to. 
We confine ourselves only to that
region of the parameter space 
where one achieves a relic density
consistent with the WMAP data. LSP annihilation 
is efficiently mediated by the pseudoscalar Higgs, with the `funnel region'
being significantly extended toward small values of $\tan\beta$ 
when compared to mSUGRA. 
Having ensured that 
the region of SUSY parameter space thus isolated is consistent with all
constraints from FCNC and CP-violation, we have proceeded to investigate
the signals of this scenario at the LHC.

Although stop-pair production is the dominant SUSY process in such 
scenarios, the stringent cuts that we choose to impose results 
in stop-cascades being suppressed. Rather, the dominant contribution to the
signal rates
turns out to be gluino pair-production.
The relatively large multiplicity
of top quarks produced in the cascades 
results in enhanced  rates
for two, three and four-lepton final states, together with missing-$E_T$ 
and hard jets. In particular, the usefulness of four-lepton
final states is highlighted through this analysis.

Based on the study of a few benchmark points (corresponding to different
gluino masses and $\tan\beta$), we find that, using the criteria chosen by us,
it is possible to probe the above scenario with an integrated luminosity
of $30 \fb^{-1}$ for gluino masses up to about 1.2 TeV. (Indeed, for 
certain significant parts of the parameter space, even $2 \fb^{-1}$ would 
be enough.) 
The reach can be 
potentially extended further once more luminosity accrues. We also 
demonstrate that it is possible to distinguish this scenario from 
an mSUGRA-one
 tuned at the same gluino mass and satisfying the
WMAP constraints. 
The usefulness of the ratios of
events in various channels is clearly elicited from our study.
Moreover, such distinction is facilitated by the effective mass distribution
of events for gluino masses on the higher side, i.e. above a TeV.
Thus, we succeed in illustrating that a multichannel analysis is not
only able to probe such nonuniversal SUSY scenarios satisfying the 
relic density constraints,
but can also highlight notable differences with a simple-minded
model based on universal SUGRA.

\section*{Acknowledgments} 

UC, DC and BM thank the organisers of the workshop 
``TeV Scale Physics and Dark Matter'', held in NORDITA, Stockholm in 
2008 for the kind hospitality received. 
The work was initiated in this workshop. 
The work of SB and BM was partially supported by funding
available from the Department of Atomic Energy, Government of India
for the Regional Centre for Accelerator-based Particle Physics,
Harish-Chandra Research Institute. SB thanks Sanjoy Biswas for 
useful discussions. DC acknowledges support from 
the Department of Science and Technology, India under 
project number SR/S2/RFHEP-05/2006.

\end{document}